\documentclass[aps,pre,preprint,groupedaddress]{revtex4}
\usepackage{graphicx} 
\usepackage{dcolumn} 
\usepackage{bm}
\newcommand{\ignore}[1]{} 
\usepackage{epsfig} 
\usepackage{color}
\usepackage{amsmath,amssymb}

\begin{document}

\title{Interacting opinion and disease dynamics in multiplex networks:
discontinuous phase transition and non-monotonic consensus times}

\author{F\'atima Vel\'asquez-Rojas} 

\author{Federico Vazquez} 

\affiliation{IFLYSIB, Instituto de F\'isica de
  L\'iquidos y Sistemas Biol\'ogicos (UNLP-CONICET), 1900 La Plata,
  Argentina}

\date{\today}

\begin{abstract}
Opinion formation and disease spreading are among the most studied
dynamical processes on complex networks.  In real societies, it is
expected that  these two processes depend on and affect each other.  
However, little is known about the  effects of opinion
dynamics over disease dynamics and vice versa, since most studies treat
them separately.  In this work we study the dynamics of the  voter
model for opinion formation intertwined with that of the contact
process for disease spreading, in a population of agents that interact
via two types of connections, social and contact.  These two
interacting dynamics take place on two layers of networks, coupled
through a fraction $q$ of links present in both networks.  The
probability that an agent updates its state depends on both, the
opinion and disease states of the interacting partner.  We find that
the opinion dynamics has striking consequences on the statistical
properties of disease spreading.  The most important is that the
smooth (continuous) 
transition from a healthy to an endemic phase observed in the contact
process, as the infection probability increases beyond a threshold,
becomes abrupt (discontinuous) in the two-layer system.  Therefore,
disregarding the effects of social dynamics on epidemics propagation
may lead to a misestimation of the real magnitude of the
spreading.  Also, an endemic-healthy 
discontinuous transition is found when the coupling $q$ overcomes a
threshold value.  Furthermore, we show that the disease dynamics delays
the opinion consensus, leading to a consensus time that varies 
non-monotonically with $q$ in a large range of the model's parameters.
A mean-field approach reveals that the coupled dynamics of opinions
and disease can be approximately described by the dynamics of the
voter model decoupled from that of the contact process, with
effective probabilities of opinion and disease transmission.
\end{abstract}

\maketitle

\section{Introduction}
\label{introduction}

The formation of opinions and the propagation of an epidemic disease
on a population of individuals are among the most studied dynamical
processes on complex networks \cite{Barrat-2008,Castellano-2009}.  The
behavior of each of these two processes has been explored
independently of one another for the last decades, and many of their
propagation properties on diverse complex topologies are well
established already (see     \cite{Castellano-2009} and
\cite{Pastor-Satorras-2015} for recent reviews on opinion formation
and epidemic spreading, respectively).  However, less attention has
been paid to a possible case scenario  where the dynamics of opinions
interact with that of the disease spreading.  In fact, it is hardly
expected that these two dynamics are isolated in real societies but
rather depend on and affect each other,  since they both run at the
same time on the same population: an individual can  transmit a
disease to a colleague while having a conversation and exchanging
ideas or opinions on a given topic.  Then, the following questions
arise:  does the dynamics of opinion formation have an impact on the
extent and prevalence of the epidemic?  Does the disease spread
facilitate the ultimate dominance of one opinion, or does it rather
hinder the consensus of opinions? 
 
In an attempt to explore these questions, we study in this article how
opinion formation and disease spreading    
processes affect each other, using two simple models as a
proxy of each process: the voter model (VM) and the contact process
(CP).  The VM was originally introduced as the simplest
system of interacting particles that can be exactly solvable in any
dimension \cite{Clifford-1973,Holley-1975,Liggett-2004}, and is one
of the most studied models for opinion consensus.  In this model,
individuals can take one of two possible positions or opinions on a
given issue, and are allowed to update them by adopting the opinion
of a randomly chosen neighbor.  The CP, on its part, has been
extensively studied to explore the spread of an infection in a system
of interacting agents \cite{Harris-1974}, where infected
agents can transmit the infection to susceptible neighbors in a
lattice \cite{Marro-1999} or a complex network \cite{Ferreira-2013},
and they can also recover at a given rate.  The CP exhibits a
continuous transition from a healthy to an endemic phase when the
infection rate exceeds a threshold value.   
To model the interaction between the two dynamics we implement the
framework of multilayer complex networks 
\cite{Domenico-2013,Boccaletti-2014,Kivela-2014} that consists of a
set of complex networks interrelated with one another, which
allows to study systems composed by many interdependent processes.  In
the present study we consider that the opinion dynamics takes place on a 
network of social relations --formed by individuals that influence
each other on a social issue, while the disease spreads on a network
of physical contacts --formed by people having daily face-to-face
contacts.  All individuals are in both layers of networks, but the
pattern of connections between them may be different in each layer.
The overlap of connections is taken as a measure of the coupling between the
two networks.  

The bilayer network system described above may represent a simple case
scenario where the social network    supports a process that involves
peer pressure, like the adoption of new behaviors or opinions, while
the contact network supports the spreading of a contagious viral
infection like flu, which is transmitted by proximity or direct
contact between individuals.  The different combinations   of
connections may correspond to different types of relationships between
two individuals.  For instance, two close   friends can have both a
contact and social tie, as they can see each other at work every day
and also interchange ideas on a political issue.  But it can also
happen that individuals are connected by only one type of tie; e.g.,
two colleagues having a contact or proximity relation because they
work in the same place but never talk about politics; or two friends
that never meet but discuss political ideas by electronic means
(phone, Facebook, Twitter, email, etc).   

Some related works on multilayer networks
\cite{Granell-2013,Granell-2014,Zuzek-2016,Halu-2013,Diakonova-2014,Diakonova-2016,Vazquez-2016}
have also explored the interrelation between two information spreading
processes.  For instance, in references
\cite{Granell-2013,Granell-2014} the authors analyzed  how the
awareness of a disease affects the epidemic spreading on a multiplex
network, by using the unaware-aware-unaware and the
susceptible-infected-susceptible cyclic models, respectively.   The
interplay between opinion formation and decision making processes was
studied in \cite{Zuzek-2016} using two interconnected networks.
Another work considered two political parties (two interacting
networks) that compete for votes in a political election
\cite{Halu-2013}.  In a recent article \cite{Diakonova-2014}, the
authors studied the dynamics  of the voter model on bilayer networks
with coevolving connections, while in \cite{Diakonova-2016} the same
authors explored whether is appropriate to reduce the dynamics of the
voter model from a two-layer multiplex network to a single layer.  A
recent work \cite{Czaplicka-2016} considers a complex threshold
dynamics that competes with a simple Susceptible-Infected-Susceptible
dynamics on two interconnected networks.  All the works listed above
explore the interplay between two social or two epidemiological
processes that are alike.  However, there is a lack of specific studies
on the interplay between opinion and disease dynamics.

In this article we show that the dynamics of opinions has striking
consequences on the disease spreading and vice versa.  The nature of
the healthy-endemic transition observed in the CP, as the infection
probability increases, is largely modified by the dynamics of the VM.
The transition changes from continuous to discontinuous when the
disease and opinions are coupled,  showing a jump in the disease
prevalence at the transition point, where the magnitude of the jump
increases with the coupling.  Also, a discontinuous transition from an
endemic to a healthy phase is found when the coupling overtakes a
threshold value.  In addition, we find that the dynamics of the CP has
important consequences in   the dynamical properties of the VM.  The
diffusion of opinions is slowed down by the disease in a non-trivial
manner as the coupling increases.  This leads to consensus times that
vary either monotonically or non-monotonically with the coupling, for
a large range of the model's parameters.  We develop a mean-field
approach to study the time evolution of macroscopic quantities, which
takes into account state correlations between neighbors in the same
network (pair approximation).  This approach reveals that the
interdependent system of opinions and disease can be thought of as two
independent systems, with external parameters that depend on the
coupling.  Specifically, the opinion dynamics can be  approximated as
the dynamics of the VM on an isolated network, with an effective
probability of opinion transmission that decreases with the coupling
and the prevalence.  Analogously, the disease spreading is
approximately described by the CP dynamics on an isolated network,
with an effective infection probability that decreases with the
coupling and the fraction of neighbors with different opinions.

The article is organized as follows.  In section \ref{model}, we
introduce the multiplex framework and the dynamics of the model on
each layer.  We present simulations results in section \ref{results}
and develop an analytical approach in section \ref{analytical}.
Finally, in section \ref{summary} we give a summary and conclusions.

\section{The Model}
\label{model}

We consider a bilayer system composed by a contact and a social
network layer of mean degree $\langle k \rangle=\mu$ and $N$   nodes
each.  These two layers are interrelated through their nodes, which
are the same in both networks, while links connecting nodes may not
necessarily be the same.  That is, both layers have the same number of
nodes $N$ and links $\mu N/2$, but the configuration of connections
can be different in each layer.  The overlap of links is measured by
the fraction $q$ ($0 \le q \le 1$) of links shared by both networks.
In our model, the extreme values $q=0$ and $q=1$ correspond to totally
uncoupled and totally coupled networks, respectively.  To build this
particular topology, we start by connecting the same pairs of nodes at
random in both networks until the number of links reaches the overlap
value $q \, \mu N/2$.  Then, the rest of the links $(1-q) \mu N/2$ are
randomly placed between nodes in each network separately, making sure
that the chosen pair of nodes in one network is not already linked in
the other network.  

\begin{figure}[t]
\begin{center}
  \includegraphics[width=10.0cm]{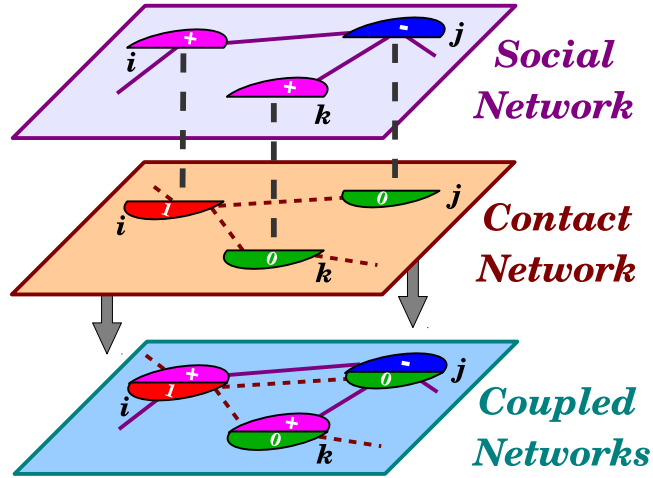}
  \caption{Schematic diagram showing a small part of a two-layer
    multiplex network.  The top layer represents a social network
    supporting the propagation of opinions, while the middle layer
    describes a network of physical contacts on which a disease
    spreads.  The bottom layer is the collapse of both layers, showing
    nodes connected by social (solid lines) and contact (dashed lines) links.  
    Node states are susceptible ($1$) and infected ($0$) in the
    contact network, and follow the contact process dynamics, while
    $+$ and $-$ states in the social network are updated according to
    the voter model dynamics.}
\label{bilayer}
\end{center}
\end{figure}

Social links in this system connect individuals that influence each
other on a given issue, while the infection is transmitted through
contact links.  In Fig.~\ref{bilayer} we illustrate the bilayer system
composed by a social and a contact network (top and middle layers),
and its representation as a single layer with two types of links
(bottom layer).  We observe in Fig.~\ref{bilayer} that nodes $i$ and
$j$ are connected by both a social and a contact link, representing
individuals that have a daily face-to-face conversation, where they
interchange opinions and also one can infect its partner.  Nodes $j$
and $k$ are only connected by a social link: they do not have
face-to-face contacts but still exchange ideas electronically or by
phone.  Nodes $i$ and $k$ are only connected by a contact link: they
have face-to-face or proximity contacts but they do not discuss and
interchange opinions about the given issue. 

To mimic the spreading of opinions and the disease we use the voter model
(VM) and the contact process (CP) on each layer, respectively.  Each
node is endowed with an opinion state $\mathcal O$ that can take two
possible values $\mathcal O=+,-$ (see top layer of Fig.~\ref{bilayer}),
and a disease state $\mathcal D=0,1$ that represents
the susceptible and infected states of an individual, respectively 
(middle layer of Fig.~\ref{bilayer}).  These two
dynamics are coupled through the opinion and disease states of nodes,
which affect each other by reducing the flow of information between
neighbors, as we describe below with a simple example.  

Let's consider a situation in which two individuals have a daily
social and physical contact because they see each other at work
and talk about politics.  On the one hand, we assume that each
individual is less influenced by its partner when she/he is sick,
because the sick partner normally stays at home or at 
hospitals, reducing physical contacts between them.  This 
makes social relations (and the interchange of opinions) less likely when at
least one of the two individuals is sick.  Thus, we consider that a
social relation 
takes place with probability $1.0$ if both social neighbors are
healthy, and with a reduced probability $p_o \le 1.0$ when one or both
are sick.  In case they have a social contact but not a physical
contact (they do not see each other but discuss ideas by electronic
means), the disease state is not supposed to affect the probability of
social interactions between them.  Therefore, the social interaction
probability is not reduced by the disease and, for simplicity, is set
to $1.0$ as in the case of healthy neighbors.  

On the other hand, we consider that physical contacts (and therefore infections)
between the two social and contact neighbors are more likely to
happen when they share the same opinion.  This is a consequence of a 
sociological mechanism called     
homophily \cite{Axelrod-1998,McPherson-2001,Vazquez-2007-2}, i. e.,
the tendency for individuals to interact with similar 
others.  The effects of homophily in the propagation of cultural
attributes in a society were studied by Robert Alxerod using an
agent-based model \cite{Axelrod-1998}, in which the probability that
two neighboring agents interact is proportional to their
cultural similarity (the number of shared attributes).
Following this idea, we assume that the contact probability between
the neighbors when they have the same opinion is higher than that when
they have different opinions.  Therefore, we set to $1.0$ the contact
probability of same-opinion neighbors and denote by $p_d \le 1.0$ the
contact probability between opposite-opinion neighbors.  Once they
have a physical contact the infection is transmitted with probability
$\beta$, leading to effective infection probabilities $\beta$ and 
$\beta \, p_d\le \beta$ in each respective case.     
In a situation where there is a contact but not a social connection
between two neighbors (they see each other but they don't talk about
politics), opinions are not expected to affect (neither increase or
decrease) the contact probability.  Therefore, this 
can be considered as an intermediate situation respect to the two
cases mentioned above, where the contact
probability should be smaller than $1.0$ but larger than $p_d$, leading to an
infection probability between $\beta$ and $\beta \, p_d$.  However, for
simplicity we assume that the contact probability in the absence of a
social relation is the same as that in homophilic relations ($1.0$), and
thus the infection probability takes the value $\beta$.  This
approximation and the one mentioned above for the social 
interaction probability have the advantage of reducing the number of
free parameters, allowing for a deeper analysis of the
model which already exhibits a very rich behavior as we shall see.    

\begin{figure}[t]
\begin{center}
  \includegraphics[width=12.5cm]{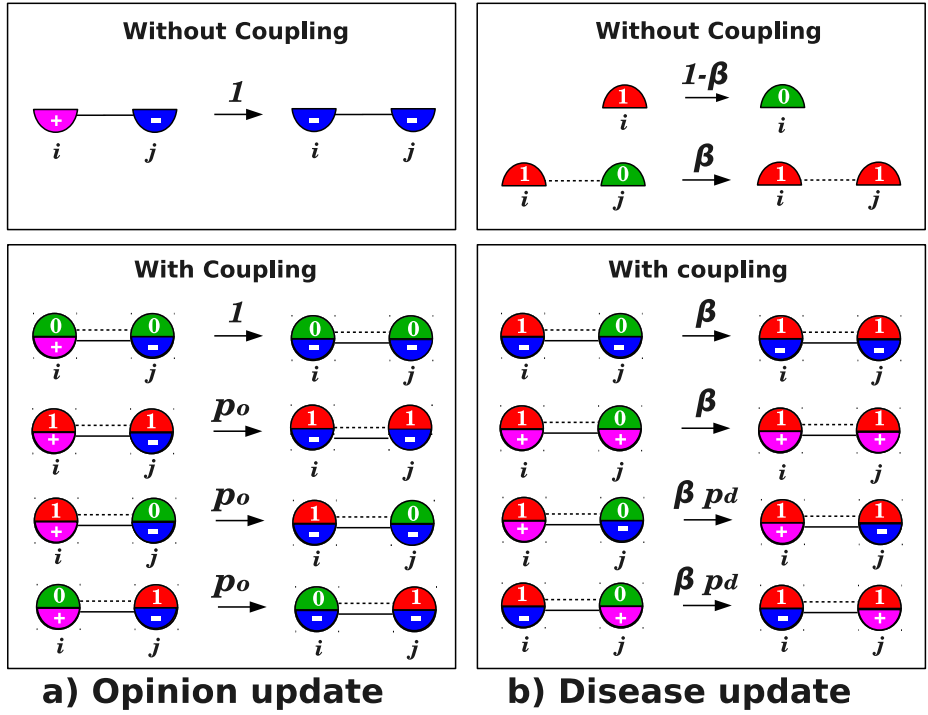}
  \caption{Update rules in the coupled opinion-disease system.  
  (a) Opinion update.  Node $i$ adopts the opinion of its neighbor $j$
  with probability $1.0$ when they are connected only by a social link
  (solid line).  When they are also connected by a contact link
  (dashed line), adoption happens with probability $1.0$ if both nodes
  are susceptible, and with probability $p_o \le 1.0$ if at least one
  node is infected.  (b) Disease update. An infected node $i$ recovers
  with probability $1-\beta$ or transmits the disease to a susceptible
  neighbor $j$ with probability $\beta$ when both nodes are only connected by a
  contact link, or when they are connected by both types of links and
  they share the same opinion.  In case they hold opposite opinions
  the transmission happens with probability $\beta \, p_d \le \beta$.}
\label{updates}
\end{center}
\end{figure}
 
We now define the dynamics of the model according to the interaction
properties discussed above.  In a single time step $\Delta t=1/N$ an
opinion and a disease update attempt take place in each network, as we
describe below (see Fig.~\ref{updates}).  

\emph{Opinion Update} [Fig.~\ref{updates}(a)]: a node $i$ with opinion
$\mathcal O_i$ and one of its neighbors $j$ with opinion $\mathcal
O_j$ are randomly chosen from the social network.  If $\mathcal O_i =
\mathcal O_j$ nothing happens.  If $\mathcal O_i \neq \mathcal O_j$,
then $i$ copies the opinion 
of $j$ ($\mathcal O_i \to \mathcal O_i = \mathcal O_j$) with
probability $p_o$ if there is a contact link between $i$ and $j$, and
at least one of the two nodes is infected ($\mathcal D_i=1$ or
$\mathcal D_j=1$).  Otherwise, i e., if there is no contact link or
$\mathcal D_i = \mathcal D_j=0$, then $i$ copies $j$'s opinion with
probability $1.0$.     

\emph{Disease update} [Fig.~\ref{updates}(b)]: a node $i$ with disease
state $\mathcal D_i$ is 
chosen at random from the contact network.  If $\mathcal D_i=0$
nothing happens.  If $\mathcal D_i=1$, then $i$ recovers with probability 
$1-\beta$ or, with the complementary probability $\beta$ node $i$ tries to
infect a randomly chosen neighbor $j$, as long as it is in the
susceptible state ($\mathcal D_j=0$).  The
infection happens ($\mathcal D_j=0 \to \mathcal D_j=1$) with
probability $p_d$ if there is a social link 
between $i$ and $j$, and $\mathcal O_i \neq \mathcal O_j$.  Otherwise,
i e., if there is no social link or $\mathcal O_i = \mathcal O_j$, then 
node $j$ is infected with probability $1.0$. 

In other words, individuals on the social layer adopt the 
opinion of their neighbors with probability $1.0$ except when they are
connected by a contact link and one of them is infected, where in this
case the opinion is adopted with a reduced probability $p_o \le 1$ [see
Fig.~\ref{updates}(a)].  The CP dynamics happens on the disease layer with an
infection probability $\beta$ between two neighbors, which is reduced to
$\beta\, p_d \le \beta$ only in the case they are attached by a social
link and they share different opinions [see Fig.~\ref{updates}(b)].

\section{Numerical Results}
\label{results}

The CP and the VM are two of the most studied dynamical processes
\cite{Liggett-2004}.  A relevant feature of the CP is the existence of a
transition from a healthy phase to an endemic phase as the infection
probability overcomes a threshold value $\beta_c$.  The healthy phase
is static, as all nodes are susceptible and infection events cannot occur.
The endemic phase is active, where each node undergoes an
infected-susceptible-infected cycle and the total number of infected
nodes fluctuates around a stationary value.  The healthy-endemic transition is continuous, and the critical value $\beta_c$
depends on the topological properties of the network
\cite{Ferreira-2013}.  For its part, the VM has been extensively used to explore
opinion consensus on different network topologies \cite{Castellano-2003,Vilone-2004,Suchecki-2005,Castellano-2005,Sood-2005,Vazquez-2008-2}.  It was found that
the diffusion properties of opinions depend on the heterogeneity of
the network.  This is reflected in the mean consensus time, which is
proportional to the ratio $\mu^2/\mu_2$ \cite{Sood-2005,Vazquez-2008-2},
where $\mu$ and $\mu_2$ are the first and second moments of the
network's degree distribution.  

The behavior described above is particular of each model on single
isolated networks.  In order to explore 
how the properties of these two 
processes are affected when they are coupled through a multiplex network, we run
extensive Monte Carlo (MC) simulations of the model described in 
section \ref{model}, using two Erd\"{o}s-R\'enyi (ER) networks of mean
degree $\langle k \rangle = \mu=10$ each.  Initially, each node in the
system is infected with probability $1/2$, and adopts either opinion
state $+$ or $-$ with equal probability $1/2$.  That is, the system
starts from a symmetric initial condition with roughly $1/4$ of 
nodes in each of the four possible opinion-infection states: 
$\left[{+ \atop 0}\right]$, $\left[{+ \atop 1}\right]$,
$\left[{- \atop 0}\right]$ and $\left[{- \atop 1}\right]$.

In the next two subsections we study separately the effects of one
dynamics over the other. 

\subsection{Effects of opinion formation on disease prevalence}
\label{opinion-disease}

\begin{figure}[t]
\begin{center}
\includegraphics[width=7.5cm]{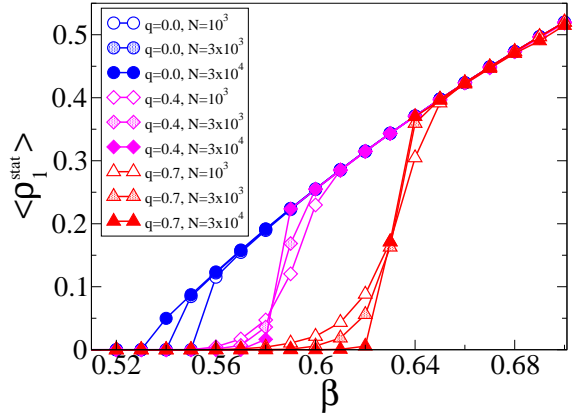}
\caption{Average stationary fraction of infected nodes 
  $\langle \rho_1^{\mbox{\tiny stat}} \rangle$ vs
  infection probability $\beta$ on two coupled ER networks of mean 
  degree $\mu=10$ and $N$ nodes each, for $p_o=p_d=0$ and coupling  
  parameters $q=0$ (circles), $q=0.4$ (diamonds) and $q=0.7$
  (triangles).  Different symbol fillings correspond to three
  different sizes
  $N=10^3, 3 \times 10^3$ and $3 \times 10^4$, as indicated in the legend. 
  The average was done over 
  $5000$ independent realizations starting from configurations
  consisting on a fraction close to $50\%$
  of infected nodes uniformly distributed over the contact network and
  $50\%$ of $+$ opinions uniformly distributed over the social network.}     
\label{rho-beta-q}
\end{center}
\end{figure}

We start the analysis of the model by describing the results related
to the effects of opinion formation on the properties of disease
spreading.  In Fig.~\ref{rho-beta-q} we show the   stationary fraction
of infected nodes averaged over many independent  realizations of the
dynamics, $\langle \rho_1^{\mbox{\tiny stat}} \rangle$, as a function
of the infection probability $\beta$.  For this first set of
simulations we used $p_o=p_d=0$, which corresponds to the extreme case
scenario where opinions cannot be transmitted across contact neighbors
(nodes connected by a contact link) that are infected, and infections
are not allowed between social neighbors (nodes connected by a social
link) with different opinions.  Different curves correspond to
different values of the coupling parameter $q$ and network size $N$,
as indicated in the legend.  We observe that, for $q=0.4$ (diamonds)
and $q=0.7$ (triangles), $\langle \rho_1^{\mbox{\tiny stat}} \rangle$
decreases smoothly with $\beta$ until a point $\beta_q^c$ that depends
on $q$, where it suddenly decays to a value close to zero.  The
sudden decrease in  $\langle \rho_1^{\mbox{\tiny stat}} \rangle$
becomes more abrupt as $N$ increases, leading to a discontinuous
change of $\langle \rho_1^{\mbox{\tiny stat}} \rangle$ at $\beta_q^c$
in the thermodynamic  limit ($N \to \infty$).  This behavior is
reminiscent of a discontinuous transition.  We also see that the jump
in $\langle \rho_1^{\mbox{\tiny stat}} \rangle$  decreases with $q$
and vanishes for the uncoupled case $q=0$, where the transition
becomes continuous, in agreement with the known  behavior of the CP on
isolated networks.  The critical point $\beta_0^c \simeq 0.53$ for
$q=0$ agrees very well with the one found in previous numerical and
analytical works \cite{Ferreira-2013}.  

These results show that the dynamics of opinions has a profound
effect on the statistical properties of disease spreading, changing the
type of phase transition in the CP from a continuous transition 
in the absence of coupling (when the two dynamics are independent)
to a discontinuous transition when the dynamics are coupled.

\begin{figure}[t]
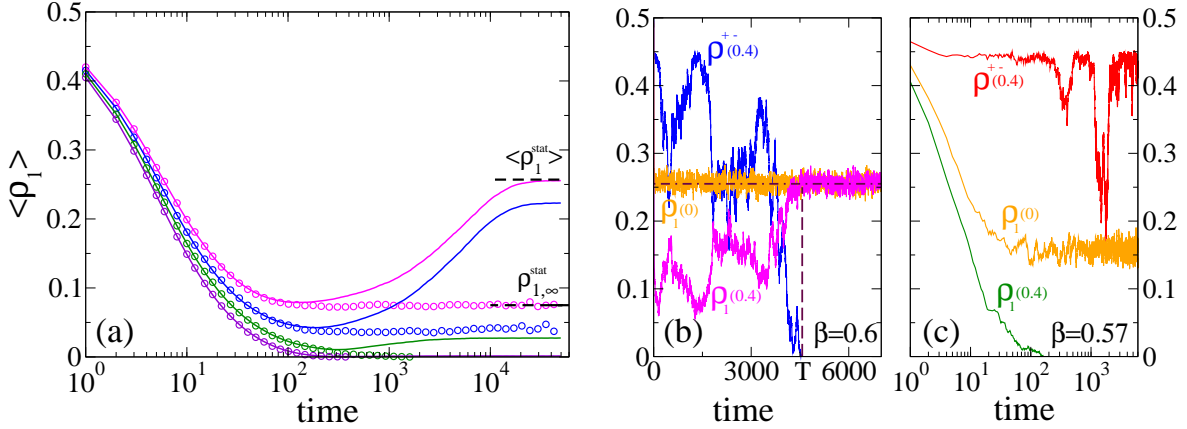

\begin{center}
\includegraphics[width=7.5cm]{Fig4a.eps} \hspace{0.3cm}
\includegraphics[width=7.5cm]{Fig4bc.eps}
\caption{(a) Time evolution of the average fraction of infected nodes
  $\langle \rho_1 \rangle$ on a bilayer system with coupling $q=0.4$.
  Solid lines correspond to networks of size $N=10^4$, while open
  circles are for networks with $N=10^6$ nodes.  Curves correspond to infection
  probabilities $\beta=0.60$, $0.59$, $0.58$ and $0.57$ (from top to
  bottom).  Horizontal dashed lines indicate the stationary values for
  $\beta=0.60$ and the two network sizes.  (b) and (c) Evolution of
  the fraction of infected nodes, $\rho_1$, and the fraction of $+-$
  social links, $\rho^{+-}$, in two distinct realizations for $q=0.4$
  and $\beta=0.6$ (b) and $\beta=0.57$ (c).  The evolution
  of $\rho_1$ is also shown for $q=0$ in both panels.}    
\label{rho-t-0.4}
\end{center}
\end{figure}

In order to achieve a deeper understanding of the nature of this
transition we studied the time evolution of the fraction of infected
nodes $\rho_1(t)$ for the case $q=0.4$, where the transition point
is $\beta_{0.4}^c \simeq 0.58$ (see Fig.~\ref{rho-beta-q}). 
Solid lines in Fig.~\ref{rho-t-0.4} correspond to results for networks
of size $N=10^4$.  As we can see, for $\beta > \beta_{0.4}^c \simeq 0.58$
the average value of $\rho_1(t)$ over many realizations, $\langle
\rho_1(t) \rangle$, varies 
non-monotonically with time and asymptotically approaches a stationary
value $\langle \rho_1^{\mbox{\tiny stat}} \rangle$ that depends on $\beta$, while 
$\langle \rho_1(t) \rangle$ decays to zero for $\beta < \beta_{0.4}^c$.
That is, this non-monotonicity in $\langle \rho_1(t) \rangle$ makes
$\langle \rho_1^{\mbox{\tiny stat}} \rangle$ jump from a value close to zero for
$\beta < \beta_{0.4}^c$  ($\langle \rho_1^{\mbox{\tiny stat}}
\rangle \simeq 0.0014$ for $\beta=0.57$) to a much larger value for 
$\beta > \beta_{0.4}^c$ ($\langle \rho_1^{\mbox{\tiny stat}} \rangle
\simeq 0.22 $ for $\beta=0.59$).    
We note that this peculiar non-monotonic temporal behavior is known to induce
discontinuous transitions in social models with multiple states and
constrains, like the Axelrod model (see for instance
\cite{Castellano-2000,Vazquez-2007-2}).  

As we explain below, the origin of the non-monotonic behavior of
$\langle \rho_1^{\mbox{\tiny stat}} \rangle$ is in the dynamic nature
of the infection probability during  each single realization, which
can take two possible values:  either the value $\beta \, p_d =0$ a
cross a contact link that overlaps with a $+-$  social link, or the
value $\beta$ otherwise (simulations correspond to $p_d=p_o=0$).  In
other words, the infectivity across a given link $i-j$ may   switch
between $0$ and $\beta$ over time, depending on the opinion states of
nodes $i$ and $j$.  This gives an average infection rate over the
entire system that fluctuates according to the evolution of the
fraction of $+-$ links, $\rho^{+-}(t)$, in one realization.  We shall
exploit this  observation in section \ref{analytical} to develop a
mean-field (MF) approach for the evolution of the system.  In panels
(b) and (c) of Fig.~\ref{rho-t-0.4} we plot $\rho_1$ and $\rho^{+-}$
in a single realization of the dynamics, for $q=0.4$ and two values of
$\beta$.  For $\beta=0.60 > \beta_{0.4}^c$ [panel (b)] we observe
that $\rho_1$ displays large variations up to a time $T \simeq 4570$
(vertical dashed line) where $\rho^{+-}$ becomes $0$, after which
$\rho_1$ fluctuates around a stationary value  $\rho_1^{\mbox{\tiny
    stat}} \simeq 0.255$  (horizontal dash-dotted line), while for
$\beta=0.57<\beta_{0.4}^c$ [panel (c)] $\rho_1$ rapidly decays to
zero, before $\rho^{+-}$ reaches zero.  When $\rho^{+-}$ becomes zero
[panel (b)] only $++$ or $--$ links remain and, therefore, the disease
dynamics behaves as the one of the standard CP with  infection
probability $\beta=0.6$ across all links, reaching the stationary
value $\rho_1^{\mbox{\tiny stat}}(q=0,\beta=0.6) \simeq 0.255$.  We
can say that after time $T$ the disease dynamics uncouples from the
opinion dynamics.  Indeed, panel (b) also shows $\rho_1$ in a single
realization on an isolated network ($q=0$) with $\beta=0.6$, where we
observe a very quick decay to a stationary value that overlaps with
the one for the coupled case $q=0.4$.  Therefore, as we can see in
Fig.~\ref{rho-beta-q}, the value of $\rho_1^{\mbox{\tiny stat}}$ in
the endemic phase of the coupled system ($\beta > \beta_{0.4}^c$) is
the same as in the uncoupled case.  Then, at the transition point
$\beta_{0.4}^c \simeq 0.58 > \beta_0^c \simeq 0.53$,
$\rho_1^{\mbox{\tiny stat}}$ jumps from the value
$\rho_1^{\mbox{\tiny stat}}(q=0,\beta_{0.4}^c) \simeq 0.22$
corresponding to the 
uncoupled system, to the small value $\rho_1^{\mbox{\tiny stat}}
\simeq 0.027$, showing a discontinuous change.  This particular
behavior of $\rho_1^{\mbox{\tiny stat}}$ is the origin of the
discontinuous transitions for $q>0$ shown in Fig.~\ref{rho-beta-q}. 

\begin{figure}[t]
\begin{center}
\includegraphics[width=7.5cm]{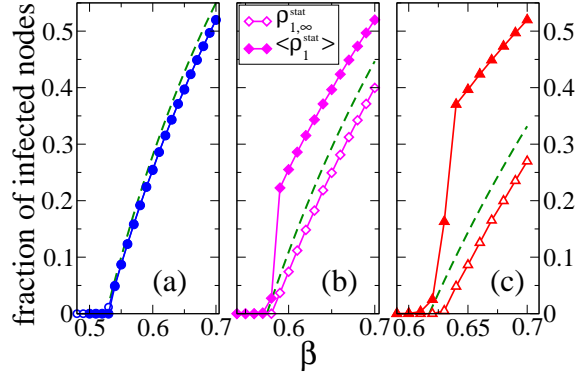}
\caption{Stationary fraction of infected nodes vs $\beta$, for
  couplings $q=0$ (a), $q=0.4$ (b) and $q=0.7$ (c).  Open symbols
  correspond to a single realization on a network of size 
  $N=10^6$, while filled symbols correspond to an average over $5000$
  realizations on networks of $N=10^4$ nodes.
  In panel (a), open symbols overlap with filled symbols.  Dashed
  curves represent the theoretical approximation from
  Eq.~(\ref{rho1stat-2}).} 
\label{rho-beta-1e6}
\end{center}
\end{figure}
     
In section \ref{analytical} we develop a MF approach that allows to
estimate the stationary fraction of infected nodes $\rho_1$ [see 
Eq.~(\ref{rho1stat-2})].  The theoretical approximation from
Eq.~(\ref{rho1stat-2}), shown as a dashed curve in each panel of
Fig.~\ref{rho-beta-1e6}, describes a continuous transition with
$\beta$,  in contrast with the discontinuity found in numerical
simulations (solid symbols).  This is because the MF approach assumes
an infinitely large system ($N=\infty$) where finite-size fluctuations
are neglected, while simulations correspond to the limit of very
large (but still finite) systems ($N \gg 1$).   Fluctuations in finite
networks ultimately drive the system to an  absorbing state in which
all nodes are susceptible ($\rho_1=0$) and have either opinion $+$ or
$-$ ($\rho^{+-}=0$), i e., an opinion consensus on a completely
healthy population.  Therefore, fluctuations play a fundamental role
in the discontinuous nature of the transition because, as previously
discussed, the stationary value of $\rho_1$ in a single realization
depends on whether $\rho^{+-}$ becomes zero before $\rho_1$ does.  
To gain a better understanding of the results obtained from the MF theory
we run simulations on very large networks.  Open circles in
Fig.~\ref{rho-t-0.4} correspond to single realizations on a network of
$N=10^6$ nodes, for the same values of $\beta$ as for networks with
$N=10^4$ nodes (solid lines).  We observe that curves for $N=10^6$
decay monotonically with time to a stationary value denoted by
$\rho_{1,\infty}^{\mbox{\tiny stat}}$ (only shown for $\beta=0.6$),
which  agrees with the minimum of the non-monotonic curves for
$N=10^4$.  We need to note that these states are not truly stationary,
in the sense that $\rho_1$ exhibits a very long plateau (outside the
shown scale) but eventually increases and reaches the same stationary
value $\langle \rho_1^{\mbox{\tiny stat}} \rangle$ of the curves for
$N=10^4$.  We have checked that the length of the plateau diverges
with $N$, and thus is infinitely large when $N=\infty$.  Therefore, we
take $\rho_{1,\infty}^{\mbox{\tiny stat}}$ as the stationary value
when $N=\infty$.  In Fig.~\ref{rho-beta-1e6} we observe that the
numerical values $\rho_{1,\infty}^{\mbox{\tiny stat}}$ (open symbols)
agree reasonable well with the theoretical approximation from
Eq.~(\ref{rho1stat-2}) (dashed curves) for the three values of $q$,
even though the agreement worsens as $q$ gets larger.  We also see
that $\rho_{1,\infty}^{\mbox{\tiny stat}}$  decays continuously as
$\beta$ decreases and becomes zero at the same value $\beta_{q}^c$ of
the transition in the thermodynamic limit corresponding to
$\rho_1^{\mbox{\tiny stat}}$ (filled symbols).  That is, the
healthy-endemic transition is continuous in an infinite system.

\begin{figure}[t]
\begin{center}
\includegraphics[width=7.5cm]{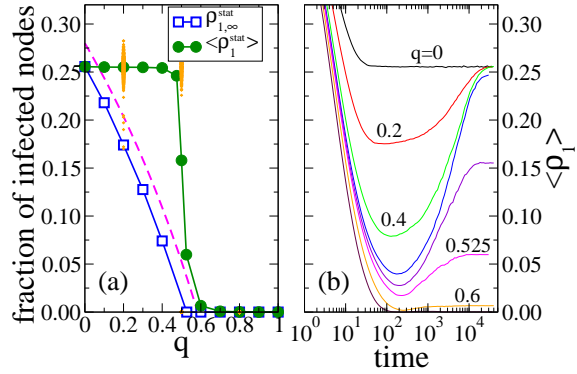}
\caption{(a) Stationary fraction of infected nodes
  $\rho_1^{\mbox{\tiny stat}}$ vs coupling $q$, for $\beta=0.6$.
  Solid circles correspond to the average of $\rho_1^{\mbox{\tiny stat}}$ 
  over $5000$ realizations on networks with $N=10^4$ nodes.  Some of
  the values of $\rho_1^{\mbox{\tiny stat}}$ in a single realization
  are shown by dots, for $q=0.2$, $0.5$ and $0.8$.  Open squares
  represent results of  
  $\rho_1^{\mbox{\tiny stat}}$ in a single realization on a network of
  size $N=10^6$.  The dashed curve is the
  theoretical approximation from Eq.~(\ref{rho1stat-2}).  (b)  Time evolution of
  $\langle \rho_1 \rangle$ for couplings $q=0, 0.2, 0.4, 0.475,
  0.5, 0.525, 0.6$ and $0.7$ (from top to bottom) on networks of size
  $N=10^4$.}    
\label{rho-q}
\end{center}
\end{figure}

Up to here we studied the response of the system when the infection
probability is varied, for a fixed coupling.  We now explore the
effects of having a varying coupling on disease prevalence.  In
Fig.~\ref{rho-q}(a) we plot $\langle \rho_1^{\mbox{\tiny stat}}
\rangle$ on two coupled  networks of $N=10^4$ nodes (circles), and
$\rho_1^{\mbox{\tiny stat}}$ in a single realization on networks of
size $N=10^6$ (squares), as a function of the  coupling $q$, for
$\beta=0.6$.  The upper curve for $N=10^4$ shows an abrupt transition
from an endemic to a healthy phase as the coupling overcomes a
threshold value $q_{0.6}^c \simeq 0.5$.  To explore this behavior in
more detail, we show with dots the value of  $\rho_1^{\mbox{\tiny
    stat}}$ in every single realization for three values of $q$.
For $q=0.2$, all dots fall around its mean value $\langle
\rho_1^{\mbox{\tiny stat}} \rangle \simeq 0.26$, while for  $q=0.8$
they are at $\rho_1^{\mbox{\tiny stat}}=0$.  At the transition point
$q_{0.6}^c$ the distribution of dots is bimodal, i e., dots are around
$\rho_1^{\mbox{\tiny stat}} \simeq 0.26$ and at $\rho_1^{\mbox{\tiny
    stat}} =0$, giving an  average value $\langle \rho_1^{\mbox{\tiny
    stat}} \rangle \simeq 0.165$.  This is an evidence of a
discontinuous transition.  The reason for this discontinuity is the
non-monotonic time evolution of  $\langle \rho_1 \rangle$ [see
  Fig.~\ref{rho-q}(b)], similarly to what happens when $\beta$ is
varied, as shown before.  The only difference with this previous
studied case is that, as $\beta$ is fixed, the stationary value of
$\rho_1$ in single realizations does not change with $q$, but is
either $\rho_1^{\mbox{\tiny stat}}=0$ or $\rho_1^{\mbox{\tiny stat}}
\simeq 0.26$, in  agreement with the binomial distribution.  The
former situation  happens in realizations where $\rho_1$ hits zero
before $\rho^{+-}$ does, while the later corresponds to realizations
where $\rho^{+-}$ becomes zero and thus the two dynamics get
uncoupled, after which $\rho_1$ reaches a stationary value similar to
$0.26$ corresponding to $q=0$.   Figure~\ref{rho-q}(a) shows that the
transition with $q$ is continuous in an infinitely large system
(squares).  One can also check that the stationary value
$\rho_1^{\mbox{\tiny stat}}$ for a given  $q$ in an infinite system
agrees with the minimum of the corresponding $\langle \rho_1 \rangle$
vs time curve of Fig.~\ref{rho-q}(b).  This behavior is akin to the
one shown in  Fig.~\ref{rho-t-0.4}(a).   
 
\begin{figure}[t]
\begin{center}
\includegraphics[width=7.5cm]{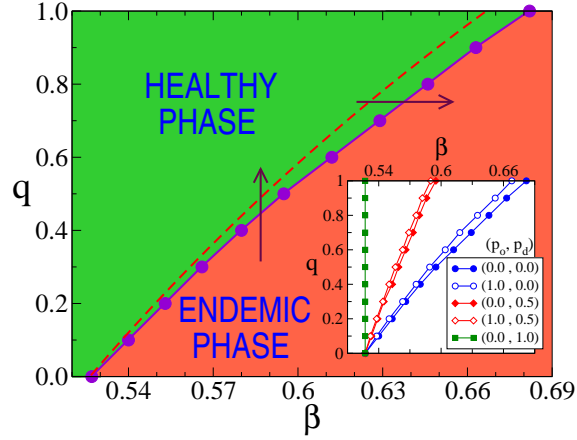}
\caption{Phase diagram of the contact process coupled to the voter
  model, showing the healthy and endemic phases in the $\beta-q$
  space, with $p_o=p_d=0$.  The dashed curve represents the analytical
  approximation of the transition line from Eq.~(\ref{qc-beta}).  The
  inset shows the transition lines for the set of values $(p_o,p_d)$
  indicated in the 
  legend.  For $p_d=1.0$ (squares), the transition is given by the vertical
  line $\beta \simeq 0.53$ for all values of $p_o$ (only $p_o=0.0$ shown).}
\label{beta-q}
\end{center}
\end{figure}

The $\beta-q$ phase diagram of Fig.~\ref{beta-q} summarizes the
results obtained in this section, on how the coupling between the
contact and social networks affects the prevalence of the disease.  By
increasing the coupling $q$ it is possible to bring an initially
uncoupled system from the endemic to the healthy phase (vertical
arrow).  Also, as the coupling increases, a larger infection
probability $\beta$ is needed to pass from the healthy to the endemic
phase (horizontal arrow). 

Finally, we reproduced the phase diagram for various values of the 
probability $p_d$ of having a successful infection across $+-$ links,
and the probability $p_o$ of opinion imitation between infected
neighbors (inset of Fig.~\ref{beta-q}).  We see that the orientation
of the transition line that separates the healthy from the endemic
phase becomes more vertical as $p_d$ increases, enlarging the endemic
phase, as we might expect.  And when $p_d=1.0$, the transition becomes
independent of the coupling $q$ and $p_o$ (the curve is the same for
all values of $p_o$).  We also observe a slight decrease of the
healthy phase when $p_o$ increases while keeping $\beta$ fixed.  As
the fraction of $+-$ links decreases faster when opinions are copied
at a higher rate, one expects an increase of the effective infection
rate and, consequently, an enlargement of the endemic phase.

\subsection{Effects of disease spreading on opinion consensus}
\label{disease-opinion}

\begin{figure}[t]
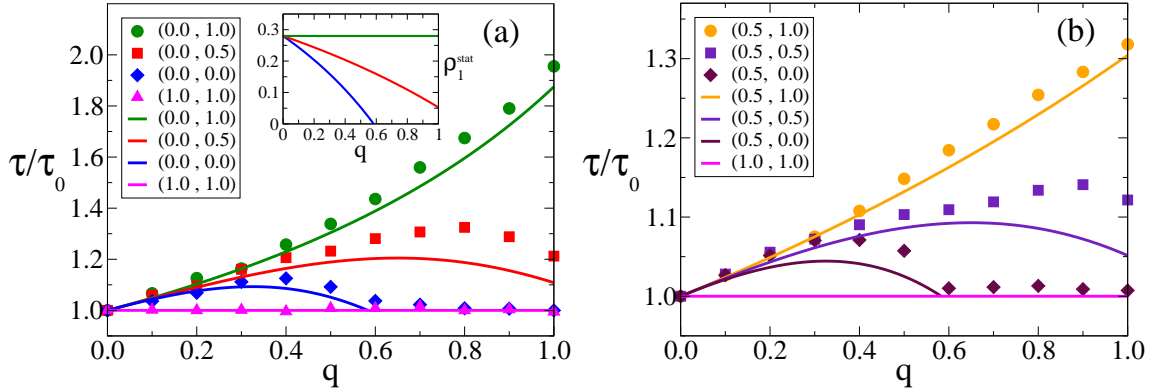

\begin{center}
\includegraphics[width=7.5cm]{Fig8a.eps}
\includegraphics[width=7.5cm]{Fig8b.eps}
\caption{(Color online) Mean time $\tau$ to reach opinion consensus on
  the social network as a function of the coupling $q$ with the
  contact network, normalized by 
  the mean consensus time in the absence of coupling $\tau_0$.  The
  infection probability 
  in the contact network is $\beta=0.6$.  Each network has $N=10^4$
  nodes and mean degree $\mu=10$.  The average was done over $5000$
  independent realizations.  Different symbols correspond to numerical 
  results for the set of values $(p_o,p_d)$ indicated in the legends
  [(a) for $p_o=0.0$ and (b) for $p_o=0.5$], while solid lines are
  the corresponding analytical approximations from Eq.~(\ref{tau-tau0}).  For
  comparison, we also show in both panels numerical data and the
  analytical curve for the uncoupled case $(p_o,p_d)=(1.0,1.0)$.
  Inset of panel (a): $\rho_1^{\mbox{\tiny stat}}$ vs $q$ from
  Eq.~(\ref{rho1stat-1}) for $p_d=1.0$, $0.5$ and $0.0$ (from top to
  bottom).}    
\label{tau-q}
\end{center}
\end{figure}

In this section we explore how the spreading of the disease affects
the dynamics of opinions.  As the transmission of opinions between
neighboring nodes is more difficult when at least one of them is sick,
we are particularly interested in studying up to what extent the
disease slows down opinion diffusion over the social network, and how
that depends on $q$, $\beta$, $p_o$ and $p_d$.  A way
to quantify this is by looking at the time to reach opinion consensus.
In Fig.~\ref{tau-q} we show how the mean consensus time $\tau$ varies
with the coupling $q$, for infection probability $\beta=0.6$ and
various values of $p_o$ and $p_d$.  For a better comparison with the voter
model on an isolated network, $\tau$ is normalized by the mean
consensus time $\tau_0$ when the networks are uncoupled ($q=0$).
Symbols correspond to MC simulations, while solid lines are the
analytical approximations from Eq.~(\ref{tau-tau0}) obtained in
section \ref{analytical}.  Here we present results for $\beta$ above
the critical point of an isolated network 
$\beta_0^c \simeq 0.53$ because for $\beta < \beta_0^c$ the effects of
disease on consensus times are negligible.  This happens because
for $\beta < \beta_0^c$ and any value of $q$ the disease quickly
disappears on the contact network and, as all nodes are susceptible,
the dynamics of opinions is decoupled from the disease dynamics,
reaching consensus in a time very similar to the one in the uncoupled
case ($\tau \simeq \tau_0$).  

We observe in Fig.~\ref{tau-q} that the $q$-dependence of $\tau$ is
quite  diverse, showing monotonic as well as non-monotonic behaviors.
This is a consequence of the competition between two different
mechanisms that directly affect opinion transmission.  One is the link
overlap between the two networks that is proportional to $q$, and the
other is the disease prevalence that decreases with $q$, as we explain
below.  The opinion transmission through a social link that overlaps
with a contact link is slowed down when at least one of the two nodes
is infected and $p_o <1$.  Therefore, the overall delay in 
opinion transmission caused by the total overlap tends to increase
with $q$, and so does $\tau$.  This effect explains the initial
monotonic increase of $\tau$ as $q$ increases from $0$, in all
curves.  However, as $q$ becomes larger a second effect becomes
important: the fraction of infected nodes decreases with $q$ [see
  inset of Fig.~\ref{tau-q}(a)], due to the coupling with the opinion
dynamics that reduces the effective infection probability as discussed
in section \ref{opinion-disease}.  Then, lower disease prevalence
translates into fewer social links affected by the disease and,
therefore, into a smaller opinion delay.  This effect tends to reduce
$\tau$ with $q$.  

With these two mechanisms at play, the shapes of curves
in Fig.~\ref{tau-q} for different values of $p_o$ and $p_d$ can be
qualitatively explained in terms of the combined effects of overlap
and prevalence.  For instance, in Fig.~\ref{tau-q}(a) we observe that
the three curves for $p_o=0.0$ have a quite different behavior.  
For $p_d=1.0$ the effect of prevalence does not vary with $q$, given
that $\rho_1^{\mbox{\tiny stat}}$ is independent of $q$ [inset of
  Fig.~\ref{tau-q}(a)]. 
Then, $\tau$ increases monotonically with $q$ as the overlap
increases.  For $p_d=0.5$ the prevalence effect increases with $q$
($\rho_1^{\mbox{\tiny stat}}$ decreases), becoming dominant for $q$
above $0.8$ when $\tau$ 
decays, and leading to a non-monotonic behavior of $\tau(q)$.
Finally, for $p_0=0.0$ we observe a non-monotonicity similar than that
of the $p_d=0.5$ curve, but with the addition that $\tau$ becomes
very similar to $\tau_0$ for all values of $q>0.6$.  This is because
$\rho_1^{\mbox{\tiny stat}}$ becomes zero above $q \simeq 0.583$ and
thus the disease has no 
effect on opinions, leading to consensus times similar to the ones
measured in isolated networks.  These behaviors for the $p_o=0.0$ case
are also observed for other values of $p_o$, as we show in
Fig.~\ref{tau-q}(b) for $p_o=0.5$.  We see that the shape of the curves
for $p_d=0.0$, 
$0.5$ and $1.0$ are analogous to the ones of Fig.~\ref{tau-q}(a) for the
corresponding values of $p_d$.  However, consensus times are smaller
for the $p_o=0.5$ case because the delay in opinion transmission is
reduced as $p_o$ increases.
             
\begin{figure}[t]
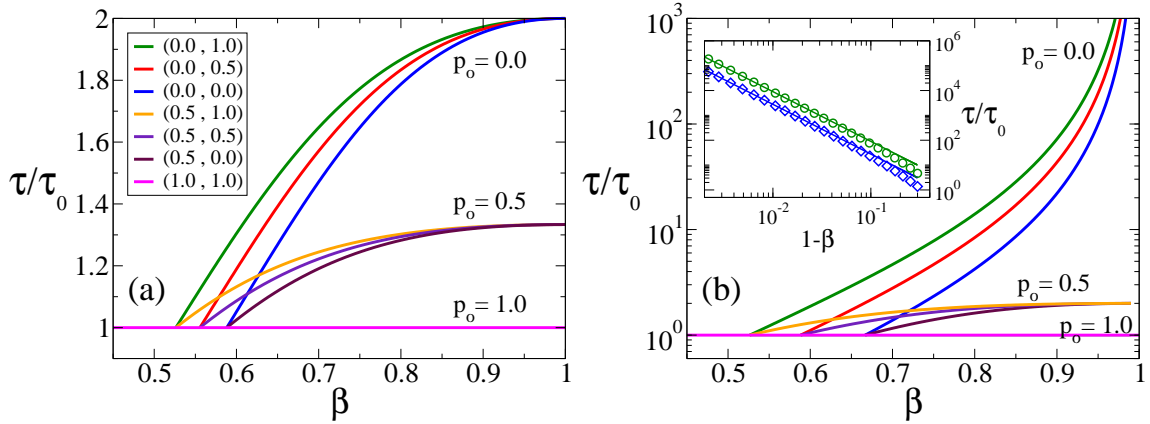

\begin{center}
\includegraphics[width=7.5cm]{Fig9a.eps}
\includegraphics[width=7.5cm]{Fig9b.eps}
\caption{(Color online)  Normalized mean consensus time $\tau/\tau_0$
  on the social 
  network as a function of the infection probability $\beta$ on the
  contact network for the same network parameters as in
  Fig.~\ref{tau-q}, coupling $q=0.5$ (a) and $q=1.0$ (b).  Curves
  correspond to Eq.~(\ref{tau-tau0}) for different values of the set $(p_o,p_d)$
  with the same color 
  code as in Fig.~\ref{tau-q}, and indicated in the legend of panel
  (a).  Curves for the sets $(1.0,0.0)$ and $(1.0,0.5)$ overlap with
  the curve for $(1.0,1.0)$ shown as the horizontal line
  $\tau/\tau_0=1$.  The inset of panel (b) shows the divergence of
  $\tau$ as $\beta$ approaches $1.0$, when $q=1.0$, $p_o=0.0$, and
  $p_d=1.0$ (circles) and $p_d=0.0$ (diamonds).  Solid lines are the
  approximations from Eq.~(\ref{tau-tau0}).}    
\label{tau-beta}
\end{center}
\end{figure}

In Fig.~\ref{tau-beta} we plot the normalized mean consensus
time $\tau/\tau_0$ as a function of the infection probability $\beta$
obtained from Eq.~(\ref{tau-tau0}).  Panels (a) and (b) correspond to couplings
$q=0.5$ and $q=1.0$, respectively.  To analyze these plots we recall
that, as explained above, consensus times increase with the level of
disease prevalence in the contact network, given that a larger disease
prevalence translates into a larger delay in opinion propagation and
in the subsequent 
consensus.  A first simple observation is that $\tau$ increases with
$\beta$ and also with $p_d$, as we expect from the fact that a larger value of
$\beta$ and $p_d$ implies a larger disease prevalence.  A second
observation is that $\tau$ decreases with the likelihood of opinion
transmission $p_o$, as explained before when we compared $\tau$ in
Fig.~\ref{tau-q}(a) with Fig.~\ref{tau-q}(b).  A third observation is
that $\tau$ approaches a value independent of $p_d$ when $\beta$
goes to $1.0$.  This is because for $\beta=1.0$ (recovery probability
equals zero) and a fixed value of $q$ and $p_o$ all nodes are infected
at the stationary state, independently on the value of $p_d$, and thus
consensus times are the same for all $p_d$.  As we see in
Fig.~\ref{tau-beta}(b), the case $q=1$ and $p_o=0$ is special because
$\tau$ diverges as $\beta$ approaches $1.0$.  This happens because in
this situation the transmission of opinions is only possible between
connected nodes that are both susceptible, which vanish in the $\beta
\to 1.0$ limit, leading to divergent consensus times.  A rough
estimation of how $\tau$ scales with $\beta$ can be obtained by
assuming that $\tau$ is proportional to the time scale associated to
the opinion transmission across two given neighboring nodes in the
social network, $i$ and $j$, with opinions $+$ and $-$, respectively.
As $q=1$, $i$ and $j$ are also neighbors in the contact network.
Starting from a situation where $i$ and $j$ are infected for high
$\beta$, the opinion transmission happens after both nodes recover.
Therefore, $\tau$ is determined by the time it takes the $1$--$1$
contact link to become a $0$--$0$ link,
which scales as $(1-\beta)^{-2}$.  In section \ref{analytical} we
derived a more accurate expression for $\tau$ that exhibits this
quadratic divergence in the $\beta \to 1$ limit, shown in the inset
of Fig.~\ref{tau-beta}(b) by solid lines.

\section{Analytical approach}
\label{analytical}

In order to gain an insight into the behavior of the two-layer system
described in section \ref{results}, we develop here a MF 
approach that allows to study the time evolution of the system in
terms of the global densities of nodes and links in different states.
We denote by $\rho^+$ and $\rho^-$ the fractions of nodes with $+$ and
$-$ opinion in the social network, respectively, and by $\rho_1$ and
$\rho_0$ the fractions of infected and susceptible nodes in the
contact network, respectively.  The fractions of social links
between $+$ and $-$ opinion nodes are denoted by $\rho^{+-}$, while
$\rho_{10}$ represents the fraction of contact links between infected and
susceptible nodes.  An analogous notation is used for $++$ and $--$
social links and for $1$--$1$ and $0$--$0$ contact links.  The fractions of
nodes $\rho^+$ and $\rho_1$ are normalized with respect to the number 
of nodes $N$ in each network, while the fractions of links $\rho^{+-}$ and
$\rho_{10}$ are normalized by the number of links $\mu N/2$ in each
network, with mean degree $\mu = \langle k \rangle$.  Given that the
number of nodes and links are conserved in each layer, the following
conservation relations hold at any time for the social layer
\begin{subequations}
\begin{alignat}{4}
\label{c11}
1&=\rho^++\rho^-, \\
\label{c12}
1&=\rho^{++}+\rho^{--}+\rho^{+-}, \\
\label{c13}
\rho^+&=\rho^{++}+\frac{1}{2}\rho^{+-}, \\
\label{c14}
\rho^-&=\rho^{--}+\frac{1}{2}\rho^{+-}, 
\end{alignat}
\label{conservation1}
\end{subequations}
and analogously for the contact layer
\begin{subequations}
\begin{alignat}{4}
\label{c21}
1&=\rho_1+\rho_0, \\
\label{c22}
1&=\rho_{11}+\rho_{00}+\rho_{10}, \\
\label{c23}
\rho_1&=\rho_{11}+\frac{1}{2}\rho_{10}, \\
\label{c24}
\rho_0&=\rho_{00}+\frac{1}{2}\rho_{10}. 
\end{alignat}
\label{conservation2}
\end{subequations}
In appendices \ref{rho+}, \ref{rho+-}, \ref{rho1} and \ref{rho10} we
develop a mean-field approach that allows to obtain the following system of
coupled differential equations for $\rho^+$, $\rho^{+-}$, $\rho_1$ and
$\rho_{10}$, respectively:
\begin{subequations}
\begin{alignat}{4}
\label{Drho+Dt}
\frac{d\rho^+}{dt}&=0, \\
\label{Drho+-Dt}
\frac{d\rho^{+-}}{dt}&=\frac{2 \, \omega \rho^{+-}}{\mu} \left[ (\mu-1) 
\left(1-\frac{\rho^{+-}}{2 \rho^+ (1-\rho^+)} \right) -1 \right], 
\end{alignat}
\label{Drhos+-Dt}
\end{subequations}
with
\begin{eqnarray}
\label{Omega}
\omega &\equiv& 1-q(1-p_o) \left(\rho_1 + \frac{\rho_{10}}{2} \right),
\end{eqnarray}
and
\begin{subequations}
\begin{alignat}{4}
\label{Drho1Dt}
\frac{d\rho_1}{dt}&= \frac{\gamma \beta \rho_{10}}{2} - (1-\beta) \rho_1, \\
\label{Drho10Dt}
\frac{d\rho_{10}}{dt}&= \frac{\gamma \beta \rho_{10}}{\mu} \left[
  (\mu-1) \left(1-\frac{\rho_{10}}{1-\rho_1} \right) -1 \right] +
2(1-\beta) (\rho_1-\rho_{10}),
\end{alignat}
\label{Drhos10Dt}
\end{subequations}
with
\begin{eqnarray}
\label{Gamma}
\gamma &\equiv& 1 - q (1-p_d) \rho^{+-}.
\end{eqnarray}
These equations represent an approximate mathematical description of
the time evolution of the model on infinitely large networks, where
finite-size fluctuations are neglected.  We note that  
Eqs.~(\ref{Drhos+-Dt}) and Eqs.~(\ref{Drhos10Dt}) are coupled through
the prefactors $\omega$ and $\gamma$, which depend on the coupling $q$
and describe the opinion and disease dynamics, respectively.  
The interested reader can find in the appendices the details of the
derivation of these equations.  For the sake of simplicity, we
assumed in the derivation that all nodes have the same  
number of neighbors $k=\mu$ chosen at random, which is equivalent to
assuming that networks are degree-regular random graphs.  However, we
expect this approximation to work well in networks with homogeneous
degree distributions like the ER networks we used in the MC
simulations.  We also implemented an homogeneous pair approximation
\cite{Vazquez-2008-2} that takes into account correlations between the
state of neighboring nodes within the same layer (intralayer pair
approximation), but neglects correlations between opinion and disease
states of both layers (interlayer annealing approximation).  That is,
we considered that the opinion state of each node is uncorrelated with
its own disease state and with its neighbors' disease states and,
conversely, that its disease state is uncorrelated with its own and its
neighbors' opinion states.  

It is instructive to analyze the structure of Eqs.~(\ref{Drhos+-Dt})
and (\ref{Drhos10Dt}).  Equations~(\ref{Drhos+-Dt}) describe the
evolution of opinions on the social layer.  From Eq.~(\ref{Drho+Dt})
we see that the fraction of $+$ nodes is conserved over time:
$\rho^+(t)=\rho^+(t=0)$ for all $t \ge 0$.   This behavior is
reminiscent of that of the VM on isolated topologies, where opinion
densities are conserved at each time step.  It seems that the  disease
dynamics is not able to break the intrinsic symmetry of opinion states
induced by the voter dynamics.    Equation~(\ref{Drho+-Dt}) for the
evolution of $\rho^{+-}$ has an extra prefactor $\omega$ compared to
the corresponding equation for the VM on isolated networks
\cite{Vazquez-2008-2}, which reveals that the disease affects the
dynamics of opinions through its prevalence level, expressed by
$\rho_1$ and $\rho_{10}$  [see Eq.~(\ref{Omega})].   As discussed in
appendix \ref{rho+}, $\omega$ can be interpreted as the ``effective
probability'' that a node $i$ adopts the opinion of a chosen social
neighbor $j$ with opposite opinion, which depends on the disease state
of both $i$ and $j$.  Within a MF approach, we can assume that
the probability that $i$ copies $j$'s opinion  depends on the disease
state of an ``average pair" of contact neighbors, and that this
probability is the same for all social neighbors.  In these terms,
$\omega$ becomes the  average copying probability over the entire
social network.  Indeed, we can check that the average value of
$\omega$ over the three possible connection and disease state
configurations of a contact pair
\begin{eqnarray*}
\mbox{copying probability} = 
\begin{cases}
1  & \mbox{with prob. $1-q$ (no contact link)}, \\ 1  &
\mbox{with prob. $q \,\rho_{00}$ ($00$ contact link)}, \\ p_o
 & \mbox{with prob. $q(1-\rho_{00})$ ($10$, $01$ or
  $11$ contact link),}
\end{cases}
\label{omega-effective}
\end{eqnarray*}
gives $\omega = 1 - q + q \left[ \rho_{00} + (1-\rho_{00}) p_o
  \right]$, which is reduced to Eq.~(\ref{Omega})  by using the
relation  $1-\rho_{00}=\rho_1+\rho_{10}/2$ that follows from
Eqs.~(\ref{c22}) and (\ref{c23}).  As we see, the overall effect of
the disease on the opinion dynamics at the  MF level is to
reduce by a factor $\omega$ the rate at which opinions change in each
node.  This effect slows down the propagation of opinions through the social
network, but it does not seem to alter the properties of the voter
dynamics. 

Equations~(\ref{Drhos10Dt}) describe the evolution of the disease on
the contact layer.  These equations have the same form as the
corresponding equations for the CP on an isolated network within the
homogeneous pair approximation \cite{Ferreira-2013}, but with a
probability of 
infection given by $\gamma \beta \le \beta$.  In analogy to the case
of $\omega$ described above, $\gamma \beta$ can be interpreted as the
``effective  probability'' that a given infected node $i$ transmits
the disease to a susceptible  neighbor $j$ on the contact layer, which
depends on the opinions of both $i$ and $j$.  Indeed, the expression
$\gamma \beta= [1 - q (1-p_d) \rho^{+-}]\beta$ from Eq.~(\ref{Gamma})
is the average infection probability on the contact network,
calculated over the three possible connection and opinion state
configurations of a social pair:    
\begin{eqnarray*}
\mbox{infection probability} = 
\begin{cases}
\beta ~~ & \mbox{with prob. $1-q$ (no social link)}, \\ \beta ~~ &
\mbox{with prob. $q (1-\rho^{+-})$ (either $++$ or $--$ social link)},
\\ \beta \, p_d ~~ & \mbox{with prob. $q \, \rho^{+-}$ ($+-$ social
  link)}.
\end{cases}
\end{eqnarray*}
Thus, our MF approach assumes that this ``effective infection
probability'' from $i$ to $j$ depends on the opinion states of an
``average pair'' of neighbors on the social layer, and that is the
same for all contact neighbors.  We can say that, at the MF level, the
disease dynamics follows the  standard CP on a single 
isolated network  with homogeneous infection probability $\gamma
\beta$ and recovery probability $1-\beta$ in each node.  Therefore,
the dynamics  of opinions has an effect on the disease dynamics
equivalent to that of an external homogeneous field acting on each
node of the contact network, reducing the probability of infection
between neighbors  by a factor $\gamma$, while keeping the same
recovery probability.  

In the next two subsections we derive analytical expressions for the
disease prevalence $\rho_1^{\mbox{\tiny stat}}$ and the mean consensus
time $\tau$, from the system of Eqs.~(\ref{Drhos+-Dt}-\ref{Gamma}).

\subsection{Disease prevalence}
\label{prevalence}

In order to study how the opinion dynamics affects the disease
prevalence, we find the fraction of infected nodes at the stationary
state $\rho_1^{\mbox{\tiny stat}}$ from Eqs.~(\ref{Drhos+-Dt}) and
(\ref{Drhos10Dt}).  We start by setting the four time derivatives to
zero, substituting $\rho_{10}$ by $2(1-\beta)\rho_1/\gamma \beta$ from
Eq.~(\ref{Drho1Dt}) into Eq.~(\ref{Drho10Dt}), and solving for
$\rho_1$.  After doing some algebra we obtain two solutions, but only
one is stable depending on the values of the parameters.  The
non-trivial solution  
\begin{eqnarray}
\rho_1^{\mbox{\tiny stat}} = \frac{\left[(\mu-1) \gamma + \mu \right]
  \beta - \mu} {\left[(\mu-1)\gamma+1 \right]\beta - 1}
\label{rho1stat}
\end{eqnarray}
corresponds to the endemic phase, where a fraction of nodes is
infected, and is stable only when the numerator  $\lambda \equiv
\left[(\mu-1) \gamma + \mu \right] \beta - \mu$ is larger than zero.
For $\lambda<0$ the stable solution is $\rho_1^{\mbox{\tiny stat}}=0$,
corresponding to the healthy phase where all nodes are susceptible,
while  $\lambda=0$ indicates the transition point between the endemic
and the healthy phase.   The expression for $\rho_1^{\mbox{\tiny
    stat}}$ from Eq.~(\ref{rho1stat}) is still not closed because it
depends on  $\rho^{+-}$, through the prefactor $\gamma$.  From
Eq.~(\ref{Drho+-Dt}) we see that the fraction of $+-$ social links
reaches a stationary value given by the expression  
\begin{equation}
\rho^{+-}_{\mbox{\tiny stat}} = \frac{2(\mu-2)}{(\mu-1)} \rho^+(0)
    [1-\rho^+(0)],
\label{rho+-stat}
\end{equation}
where we used $\rho^+=\rho^+(0)$ given that $\rho^+$ remains constant
over time, as mentioned before.  We notice that
$\omega$ does not affect the stationary value
of $\rho^{+-}$, which remains the same as in the original VM
\cite{Vazquez-2008-2}.  For a symmetric initial condition on
the social layer ($\rho^+(0)=1/2$), as the one used in the
simulations, we have $\rho^{+-}_{\mbox{\tiny stat}}=(\mu-2)/[2(\mu-1)]$.
Replacing this last   expression for $\rho^{+-}_{\mbox{\tiny stat}}$
in Eq.~(\ref{Gamma}) we obtain the following expression for $\gamma$:
\begin{equation}
\gamma = 1-\frac{q(1-p_d)(\mu-2)}{2(\mu-1)}.
\label{Gamma-1}
\end{equation}
Finally, plugging Eq.~(\ref{Gamma-1}) into Eq.~(\ref{rho1stat}) we
arrive to the following approximate expression for the stationary
fraction of infected nodes in the endemic phase:  
\begin{equation} 
\rho_1^{\mbox{\tiny stat}} = \frac{\left[ 2(2 \mu - 1) - q(1-p_d)(\mu-2)
    \right] \beta - 2 \mu}
{\left[ 2 \mu - q(1-p_d)(\mu-2) \right]\beta - 2}.
\label{rho1stat-1}
\end{equation}
For a network of mean degree $\mu=10$ and $p_d=0$,
Eq.~(\ref{rho1stat-1}) is reduced to the simple expression
\begin{equation} 
\rho_1^{\mbox{\tiny stat}} = \frac{\left( 19 - 4 q  \right) \beta
  -10}{\left( 10 - 4 q \right) \beta -1},
\label{rho1stat-2}
\end{equation}
which is plotted in Figs.~\ref{rho-beta-1e6} and \ref{rho-q} (dashed
curves).  As the MF theory is meant to work for infinitely large systems,
we also plot for comparison the numerical results
obtained from simulations for very large networks (open symbols).  We
observe that, 
in all cases, the estimated theoretical value of the fraction of
infected nodes from Eq.~(\ref{rho1stat-2}) is larger than that from
simulations.  As we  explain below, this due to the fact that
correlations between opinion and disease states are neglected by the 
MF approach.    We first notice that an infection event $0 \to
1$ between two neighbors connected by a social and a contact link is
only possible when the states of nodes are  $\left[{+ \atop 1}\right]$
and $\left[{+ \atop 0}\right]$  (a $\left[{+ + \atop 1 \; 0}\right]$
pair) or   $\left[{- \atop 1}\right]$ and $\left[{- \atop 0}\right]$
( a $\left[{- - \atop 1 \; 0}\right]$ pair), because $p_d=0$ in
 Figs.~\ref{rho-beta-1e6} and \ref{rho-q}.  Then, it is expected that
 $++$ social links are negatively 
correlated with $10$ contact links and positively correlated with $11$
and $00$ contact links, given that same-opinion neighbors tend to
infect each other and thus, at a given time, they are more likely to
be either both infected or both susceptible.  However, the theoretical
approximation assumes that $++$ social links are  uncorrelated with
$10$ contact links (see appendix \ref{rho+}) and, therefore, the
estimated probability of finding a $\left[{+ + \atop 1 \; 0}\right]$
pair is larger than that obtained when negative correlations are
considered.  The same conclusion also holds for $\left[{- - \atop 1 \;
    0}\right]$ pairs.  This leads to a theoretical overestimation of
the number of     $\left[{+ + \atop 1 \; 0}\right]$ and $\left[{- -
    \atop 1 \; 0}\right]$ pairs and, consequently, to a larger rate of
infections which increases the disease prevalence respect to numerical
results, as we see in Figs.~\ref{rho-beta-1e6} and \ref{rho-q}.

Figure~\ref{rho-beta-1e6} shows that  $\rho_1^{\mbox{\tiny stat}}$
form Eq.~(\ref{rho1stat-2}) continuously decreases and vanishes as
$\beta$ decreases beyond a threshold value, as it happens in the
standard CP.  This shows that the transition to the healthy state is
continuous within the MF approach, which assumes that the system is
infinitely large.  In Fig.~\ref{rho-q} we see that
$\rho_1^{\mbox{\tiny stat}}$ decreases  with $q$, reducing the
prevalence and inducing a transition to the healthy phase.  That is,
Eq.~(\ref{rho1stat-1}) predicts a   healthy-endemic continuous
transition as $\beta$ and $q$ are varied, which happens at the point
where $\rho_1^{\mbox{\tiny stat}}$ vanishes, leading to the relation  
\begin{equation} 
\left[ 2(2 \mu - 1) - q_c (1-p_d)(\mu-2) \right] \beta_c - 2 \mu = 0.
\label{transition}
\end{equation}
The transition line 
\begin{equation} 
q_c = \frac{19 \beta_c - 10}{4 (1-p_d) \beta_c}
\label{qc-beta}
\end{equation} 
obtained from Eq.~(\ref{transition}) for $\mu=10$ is plotted in 
Fig.~\ref{beta-q} for $p_d=0$ (dashed curve).   We can see that the
agreement with simulations is good for small values of the coupling $q$, but
discrepancies arise as $q$ increases, where the theoretical prediction
from Eq.~(\ref{qc-beta}) overestimates numerical values.  Another
simple observation that follows from Eq.~(\ref{qc-beta}) is that for 
$\beta>10/[19-4(1-p_d)]$ we obtain the nonphysical value $q_c>1$.  This
means that, in the network model, it is possible to induce a
transition by increasing the coupling only when $\beta$ is lower than 
a given value, as we see in Fig.~\ref{beta-q} for $\beta<0.68$.  

As a final remark we stress that the transitions within this MF 
approach are continuous, in agreement with simulations in very large
networks.  This is so because Eqs.~(\ref{Drhos+-Dt}) and
(\ref{Drhos10Dt}) correspond to an infinite system where
finite-size fluctuations are neglected.

\subsection{Opinion consensus times}
\label{consensus}

In this section we study the quantitative effects of the disease on
the time to reach opinion consensus.  For that, we find an analytical
estimation of the mean consensus time $\tau$ as a function of the
model parameters.  

As mentioned in section \ref{analytical}, in infinitely large systems
$\rho^+$ remains constant over time  [see Eq.~(\ref{Drho+Dt})].
However, in finite systems $\rho^+$ fluctuates until it reaches either
value $\rho^+=1$ ($+$ consensus) or $\rho^+=0$ ($-$ consensus), with both
configurations characterized by the absence of $+-$ social links
($\rho^{+-}=0$).  A typical evolution of $\rho^{+-}$ towards the
absorbing state can be seen in Fig.~\ref{rho-t-0.4} (b) for $q=0.4$ on
networks with $N=10^4$ nodes. 
That is, consensus is eventually achieved in finite systems due to the
stochastic nature of the opinion dynamics, which leads the social
network to a state where all nodes share the same opinion.  In a
single opinion update $\rho^+$ may increase or decrease by $1/N$ with
the same probability $\omega \rho^{+-}/2$,  calculated as the
probability $\rho^{+-}/2$ that a node and an opposite-opinion neighbor
are selected at random, times the probability $\omega$ of opinion
adoption.  Therefore, the stochastic dynamics of the VM can be studied
by mapping $\rho^+$ into the position of a symmetric one-dimensional
random walker on the interval $[0,1]$, with a jumping probability
proportional to $\omega \rho^{+-}_{\mbox{\tiny stat}}/2$  and a step
length of  $1/N$.  Starting from a symmetric configuration with $N/2$
nodes with $+$ opinion ($\rho^+(0)=1/2$), the walker reaches either
absorbing point $\rho^+=1$ or $\rho^+=0$ in an average number of steps
that scales as $N^2$.  Then, given that the walker makes a single step
in an average number of attempts that scales as $1/\omega
\rho^{+-}_{\mbox{\tiny stat}}$, and that the time increases by $1/N$
in each attempt, we find that the mean consensus time scales as  
\begin{equation}
\tau \sim \frac{N}{\omega \rho^{+-}_{\mbox{\tiny stat}}}. 
\end{equation}
As we see in Eq.~(\ref{rho+-stat}), $\rho^{+-}_{\mbox{\tiny stat}}$ is
independent of the disease prevalence $\rho_1$ and, therefore, 
the prevalence affects $\tau$ only through the effective copying 
probability $\omega$, 
which sets the time scale associated to opinion updates.  From
Eq.~(\ref{Omega}) we see that $\omega$ equals $1.0$ when the layers
are uncoupled ($q=0$) or when $p_o=1.0$, and thus the dynamics of
opinions is exactly the same as that of the original VM.  However,
$\omega$ is smaller than $1.0$ in the presence of coupling ($q>0$) and
$p_o<1.0$, and thus the evolution of the dynamics is ``slowed down"
--in average-- by a factor $1/\omega >1.0$, given that opinions are
copied at a rate that is $\omega$ times smaller than in the uncoupled
case.  As a consequence,  $\tau$ increases by a factor $1/\omega$
respect to the mean consensus time in the uncoupled case  $\tau_0 =
\tau(q=0) \sim N/\rho^{+-}_{\mbox{\tiny stat}}$, that is
\begin{equation}     
\frac{\tau}{\tau_0} \simeq \frac{1}{\omega}.  
\end{equation}
To obtain a complete expression for the ratio $\tau/\tau_0$ as a function of the
model's parameters we express $\omega$ in terms of 
$\rho_1^{\mbox{\tiny stat}}$, by substituting into Eq.~(\ref{Omega}) the 
stationary value of $\rho_{10}$ that follows from Eq.~(\ref{Drho1Dt}), 
$\rho_{10}^{\mbox{\tiny stat}}=2(1-\beta)\rho_1^{\mbox{\tiny stat}}/\gamma \beta$.
This leads to 
\begin{equation} 
\omega=1-q(1-p_o)\left(1+\frac{1-\beta}{\gamma \beta} \right) 
\rho_1^{\mbox{\tiny stat}}.
\label{Omega-1}   
\end{equation}
In the healthy phase is $\rho_1^{\mbox{\tiny stat}}=0$, thus
$\omega=1.0$ and $\tau=\tau_0$.  In this case, the theory predicts
that the disease has no effect on the time to consensus because there
are no infected nodes that can affect the opinion dynamics.  However,
having a value $\rho_1^{\mbox{\tiny stat}}>0$ of infected nodes 
in the endemic phase has the effect of reducing $\omega$ or,
equivalently, increasing $\tau$ respect to $\tau_0$.  Plugging
into Eq.~(\ref{Omega-1}) the expressions for $\gamma$ and
$\rho_1^{\mbox{\tiny stat}}$ from Eqs.~(\ref{Gamma-1}) and
(\ref{rho1stat-1}), respectively, and reordering some terms, we obtain
the following expression that relates $\tau$ and $\tau_0$ in the
endemic phase:
\begin{equation}  
\frac{\tau}{\tau_0} \simeq  
\left[ 1-\frac{q(1-p_o)\big\{ 2(\mu-1)-q(1-p_d)(\mu-2)\beta \big\} 
\big\{ \left[2(2\mu-1)-q(1-p_d)(\mu-2)\right]\beta -2\mu \big\}}
{\beta \big\{ 2(\mu-1)-q(1-p_d)(\mu-2) \big\} 
\big\{ \left[2\mu-q(1-p_d)(\mu-2)\right]\beta -2 \big\}} \right]^{-1}.
\label{tau-tau0}
\end{equation}
In Fig.~\ref{tau-q} we plot in solid lines the ratio $\tau/\tau_0$ vs $q$
from Eq.~(\ref{tau-tau0}) for $\mu=10$. 

We observe that the theoretical values of $\tau/\tau_0$ are smaller
than those obtained from numerical simulations (symbols) for all
combinations of $p_o$ and $p_d$ shown.  A possible explanation of
these discrepancies can be given by analyzing how correlations affect
the estimated number of different types of connected nodes, as we have
done in section \ref{prevalence} for disease prevalence.  If we take
the $p_0=0$ case, we see that an opinion change due to an interaction
between two social and contact neighbors happens only if both nodes
are susceptible, that is, when they have states  $\left[{+ \atop
    0}\right]$ and $\left[{- \atop 0}\right]$.  Then, as the theory
assumes that $+-$ social links and $00$ contact links are uncorrelated
(appendix \ref{rho+}), the estimated probability of finding a
$\left[{+ - \atop 0 \; 0}\right]$ pair is larger than that in
simulations, given that $+-$ social links are expected to be
negatively correlated  with $00$ contact links.  This negative
correlation is due to the fact that susceptible neighbors tend to
align their opinions and, therefore, they are more likely to be in the
same opinion state at a given time.  This leads to an overestimation
of the number of  $\left[{+ - \atop 0 \; 0}\right]$ pairs and,
therefore, to a larger rate of opinion transmission.  This has the
overall effect of speeding up consensus, decreasing the theoretically
estimated mean time to reach consensus respect to the mean consensus
time measured in simulations, as we see in Fig.~\ref{tau-q}.        

Even though discrepancies
with numerical results increase with the coupling $q$, the
analytic expression~(\ref{tau-tau0}) is able to capture the different 
qualitative behavior of the consensus time for several combinations
of $p_o$ and $p_d$, as we describe below.  For low values of $p_d$,
there is a transition to the healthy phase when $q$ overcomes a value
$q_c <1$ given by Eq.~(\ref{transition}) and, therefore, $\tau=\tau_0$
for all $q>q_c$ [see $p_d=0$ curves in the main plot and the inset of 
Fig.~\ref{tau-q}(a)].
As a consequence, $\tau/\tau_0$ exhibits a non-monotonic behavior
with $q$, as we described in section \ref{disease-opinion}.  For
higher values of $p_d$, the transition to the healthy phase does not
happen for the physical values $q \le 1$ used in the model's
simulations, given that $q_c>1$ from Eq.~(\ref{transition}).  In this
case, $\tau$ may either increase monotonically with $q$ for large 
$p_d$ values (see $p_d=1.0$ curves), or have a maximum at some intermediate
value for medium $p_d$ values (see $p_d=0.5$ curves).
As explained in section \ref{disease-opinion}, the non-monotonicity
is a consequence of the competition between the level of link overlap among the
two layers --which increases with $q$-- and the disease prevalence
--which decreases with $q$.  This competition can be seen
quantitatively in Eq.~(\ref{Omega-1}) for $\omega$, which has three factors that
depend on $q$ and affect $\tau$.  Besides the factor proportional to
$q$, the factor $1/\gamma$ also increases with $q$, as seen from
Eq.~(\ref{Gamma-1}).  But these two factors are balanced by
$\rho_1^{\mbox{\tiny stat}}$, which decreases with $q$.  

An interesting case is the one for full coupling $q=1.0$ and $p_o=0$,
because $\tau$ from Eq.~(\ref{tau-tau0}) diverges as $\beta$
approaches $1.0$.  This happens in the model because when $\beta=1.0$
once a node becomes infected it remains infected forever.  Then, once
all nodes become infected the opinion dynamics stops, as infected
neighboring nodes cannot interchange opinions, and thus the social layer
freezes in a mixed state of $+$ and $-$ opinions and consensus is
never achieved.  By doing a Taylor series expansion of expression
(\ref{tau-tau0}) up to to second order in the small parameter
$\epsilon = 1-\beta \ll 1$ we obtain, after some algebra,
\begin{equation}
\frac{\tau}{\tau_0} \simeq
\frac{\left[9-4(1-p_d)\right]^2}{90(1-\beta)^2},        
\label{tau-tau0-1}
\end{equation}
where we used $\mu=10$.  Equation~(\ref{tau-tau0-1}) shows that $\tau$
diverges as $(1-\beta)^{-2}$ in the $\beta \to 1.0$ limit, as shown in
the inset of Fig.~\ref{tau-beta}(b).  For $\beta=1.0$ and $p_o=0$, we can check
from Eq.~(\ref{tau-tau0}) that $\tau/\tau_0 \simeq 1/(1-q)$, which
shows the divergence of $\tau$ as the system approaches the fully
coupled state $q=1.0$.

\section{Summary and Conclusions}
\label{summary}

We proposed a bilayer network model to explore the interplay between
the dynamics of opinion formation and disease spreading in a
population of individuals.  We used the voter model and the contact process to 
simulate the opinion and the disease dynamics running on a social and
contact network, respectively.  These two networks share the same
nodes and they are coupled by a fraction $q$ of links in common.  We
showed that, when the networks are coupled, the opinion dynamics can
dramatically change the statistical properties of the disease
spreading, which in turn modifies the properties of the propagation
of opinions, as compared to the case of isolated networks.

The VM dynamics is able to change the order of the healthy-endemic
phase transition observed in the CP as the infection probability $\beta$
exceeds a threshold value $\beta_c$, from a continuous transition for the
uncoupled case to a discontinuous transition when the coupling $q$ is
larger than zero.  The magnitude of the change in the disease
prevalence at the transition point $\beta_c$ increases with $q$.
The discontinuity is associated with the
non-monotonic time evolution of the fraction of infected nodes.  This
non-monotonicity is as a consequence of the time-varying nature of the
effective infection probability, which varies over time according to
the stochastic evolution of the fraction of $+-$ social links.  The system also 
exhibits a discontinuous transition from an endemic to a healthy
phase when the coupling overcomes a value $q_c$, for a fixed value of
$\beta$.  The origin of this discontinuity is the same as that of the
discontinuous transition with $\beta$, that is, the non-monotonicity
in the time evolution of the fraction of infected nodes.  We also
obtained a phase diagram in the $\beta-q$ space showing the healthy
and endemic phases for different values of the probabilities $p_d$ and
$p_o$.  In all cases, we observed that the transition point $\beta_c$
increases with $q$.

We need to mention that changes in the order of topological and dynamical
transitions were already observed in multilayer networks
\cite{Parshani-2010,Hu-2011,Gao-2012,Zhou-2014,Radicchi-2013,Radicchi-2014,Zhao-2013}.
In real populations, the implications of having continuous in contrast
to discontinuous 
transitions are very different.  Indeed, starting from a hypothetical
situation that consists on a population of individuals with an
infection rate just below the critical value (in the healthy phase), a
small increment in $\beta$ would lead to a small number of infected
individuals in the former case, but a large number of infections 
in the later case.  Therefore, disregarding the effects of social
dynamics on epidemics propagation could lead to an underestimation of
the real magnitude of the spreading.  

We developed a mean-field approach that allowed to estimate with
reasonable precision the healthy-endemic transition line
$(\beta_c,q_c)$ as a function of the model's parameters.  This approach
reveals that the disease dynamics is 
equivalent to that of the standard CP on an isolated network, with an
effective infection probability that is constant over time and that
decreases with the coupling and the stationary fraction of $+-$ social
links, for a fixed value of $\beta$.  This means that, at the
mean-field level, the overall effect of the VM on the CP is to
decrease the effective infection probability as the coupling
increases.  Therefore, as $q$ increases, a larger value of $\beta$ is
needed to bring the system to the endemic phase, leading to an
increase of the transition point $\beta_c$ with $q$.  

On its part, the CP dynamics has the overall effect of slowing down
the propagation of opinions, delaying the process of opinion consensus
compared to the one observed in an isolated network.  The MF approach
reveals that the opinion dynamics corresponds to that of the standard
VM model on an isolated network, with a probability of opinion
transmission that decreases with $q$ and the disease prevalence.
Depending on the parameters values, the mean consensus time $\tau$ can
show a monotonic increase with $q$, as well as a non-monotonic behavior.
An insight on these results was given by the MF approach,
which allowed to obtain an approximate mathematical expression that
relates $\tau$ with the parameters.  This approach shows that the
behavior of $\tau$ with $q$ is the result of two different mechanisms
at play: the overlap of social and contact links that tends to increase
$\tau$ with $q$, which is counterbalanced by the fraction of infected nodes that
tends to decrease $\tau$ with $q$.  Therefore, the non-trivial dependence of
$\tau$ with $q$ is a consequence of the competition between these two
mechanisms. 

It is interesting to note that, despite the nontrivial interplay
between the CP and the VM, the coupled interdependent system of
opinions and disease can be roughly seen as two systems
that evolve independently of one another, where each system has an 
effective parameter that depends on the other dynamics and the
coupling.  Specifically, the opinion dynamics corresponds to that of
the VM with an effective opinion transmission probability that
decreases with the disease prevalence and the coupling, while the disease
spreading is well described by the dynamics of the CP with an
effective infection probability that decreases with the fraction of
$+-$ social neighbors and the coupling.  However, this is only an approximation
that comes from the MF analysis, which neglects correlations between
opinion and disease states.       

The results presented in this article correspond to a particular
initial state that consists on even fractions of $+$ and $-$ opinion
states and even fractions of infected and susceptible states, uniformly
distributed over the networks.  As a future work, it might be worth
studying the behavior of the system under different initial
conditions, and with uneven fractions of opinion and disease states.
For example, one can simulate a population with initial polarized
opinions based on the disease, by correlating the opinion of each node with its
disease state (for instance by infecting all nodes with $-$
opinion and leaving all $+$ opinion nodes in the healthy state).
Finally, it would be interesting to study the behavior of the present
model under different update rules.  For 
instance, we have checked a simple rule in which the connection
condition --connected or disconnected-- between two nodes in one layer 
is not taken into account for the update in the other layer.  This is
an ongoing work with some preliminary results that suggest that the
critical behavior of this new model is quite different from that of the
original model.

\section*{ACKNOWLEDGMENTS}

We thank Gabriel Baglietto and Didier Vega-Oliveros for helpful comments
on the manuscript.  We acknowledge financial support from CONICET (PIP
0443/2014).

\appendix 

\section{Derivation of the rate equation for $\rho^+$}
\label{rho+}

We denote by $\big[ \mathcal{O \atop D} \big]$ the state of a given
node, where $\mathcal O=+,-$ and $\mathcal D=1,0$ are its opinion and
disease states, respectively.  Thus, there are four possible node states: 
$\big[ {+ \atop 0} \big]$, $\big[ {+ \atop 1} \big]$, 
$\big[ {- \atop 0} \big]$ and $\big[ {- \atop 1} \big]$.  In a single
time step of the dynamics, the transitions from state  
$\big[ {\mathcal{-\atop D}} \big]$ to state $\big[ {\mathcal{+ \atop
      D}} \big]$ when a node switches opinion from $-$ to $+$ lead to
a gain of $1/N$ in $\rho^+$, while the transitions $\big[ {\mathcal{+
      \atop D}} \big] \to \big[ {\mathcal{- \atop D}} \big]$ when
there is a $- \to +$ opinion change lead to a loss of $1/N$ in $\rho^+$.  
Considering these four possible transitions, the average
change of $\rho^+$ in a single time step of time interval $\Delta
t=1/N$ is described by the rate equation
\begin{eqnarray}
\frac{d \rho^+}{dt} &=& 
\frac{d \rho^+}{dt}\Big|_{-\to+} + 
\frac{d \rho^+}{dt}\Big|_{+\to-} \nonumber \\ 
&=&\frac{1}{1/N} \left[ 
\Delta \rho^+\big|_{{- \atop 0} \to {+ \atop 0}} +   
\Delta \rho^+\big|_{{- \atop 1} \to {+ \atop 1}} +  
\Delta \rho^+\big|_{{+ \atop 0} \to {- \atop 0}} +  
\Delta \rho^+\big|_{{+ \atop 1} \to {- \atop 1}} \right],
\label{drho+dt}
\end{eqnarray}
where for instance the term $\Delta \rho^+\big|_{{- \atop 0} \to {+ \atop 0}}$
represents the average change of $\rho^+$ in a time step due to 
$\big[ {- \atop 0} \big] \to \big[ {+ \atop 0} \big]$ transitions.  In
turn, $\Delta \rho^+\big|_{{- \atop 0} \to {+ \atop 0}}$
has four contributions corresponding to the different social interactions
that lead to the $\big[ {- \atop 0} \big] \to \big[ {+ \atop 0} \big]$ 
transition.  Thus, we can write
\begin{equation}
\Delta \rho^+\big|_{{- \atop 0} \to {+ \atop 0}} =    
\Delta \rho^+\Big|_{{\overline{- +} \atop 0 \; 0} \to {\overline{+ +}
    \atop 0 \; 0} } +    
\Delta \rho^+\Big|_{{\overline{- +} \atop  \overline{0 \; 0}} \to
  {\overline{+ +} \atop \overline{0 \; 0}}} +
\Delta \rho^+\Big|_{{\overline{- +} \atop 0 \; 1} \to {\overline{+ +}
    \atop 0 \; 1} } +    
\Delta \rho^+\Big|_{{\overline{- +} \atop  \overline{0 \; 1}} \to
  {\overline{+ +} \atop \overline{0 \; 1}}},
\label{Deltarho+1}
\end{equation}
and similarly for the $\big[ {- \atop 1} \big] \to \big[ {+ \atop 1} \big]$ 
transition corresponding to the second term in Eq.~(\ref{drho+dt})
\begin{equation}
\Delta \rho^+\big|_{{- \atop 1} \to {+ \atop 1}} =    
\Delta \rho^+\Big|_{{\overline{- +} \atop 1 \, 0} \to {\overline{+ +}
    \atop 1 \, 0} } +    
\Delta \rho^+\Big|_{{\overline{- +} \atop  \overline{1 \, 0}} \to
  {\overline{+ +} \atop \overline{1 \, 0} }} +
\Delta \rho^+\Big|_{{\overline{- +} \atop 1 \, 1} \to {\overline{+ +}
    \atop 1 \, 1} } +    
\Delta \rho^+\Big|_{{\overline{- +} \atop  \overline{1 \, 1}} \to
  {\overline{+ +} \atop \overline{1 \, 1} }}.
\label{Deltarho+2}
\end{equation}
Third and fourth terms in Eq.~(\ref{drho+dt}) are obtained by
interchanging symbols $+$ and $-$ in Eqs.~(\ref{Deltarho+1}) and 
(\ref{Deltarho+2}), respectively, due to the symmetry between $+$ and $-$
opinion states.  We notice that disease states remain the same after
the interactions, as only a change in the social layer can lead to a change
in $\rho^+$.  The first term in Eq.~(\ref{Deltarho+1}) represents 
the average change in $\rho^+$ due to interactions in which a node $i$ in state 
$\big[ {- \atop 0} \big]$ copies the opinion of one its social
neighbors $j$ in state $\big[ {+ \atop 0} \big]$, changing the state of
$i$ to $\big[ {+ \atop 0} \big]$.  This interaction is
schematically represented by the symbol
$\left[\mbox{\footnotesize ${{{\overline{-+}} \atop {0 \;0}}}$}\right]$, 
where the horizontal 
line over the opinion symbols describes a social
link between $i$ and $j$.  In the same way, the symbol 
$\left[\mbox{\footnotesize ${{{\overline{-+}}\atop\overline{0 \;0}}}$}\right]$
represents an interaction 
between a $\big[ {- \atop 0} \big]$ node and a neighboring 
$\big[ {+ \atop 0} \big]$ node connected by both a social and a contact
link that are indicated by horizontal lines on top of the respective
symbols.  The second, third and fourth terms in Eq.~(\ref{Deltarho+1}) 
describe, respectively, the transitions due to an interaction of node
$i$ with a $\big[ {+ \atop 0} \big]$ social/contact neighbor, a 
$\big[ {+ \atop 1} \big]$ social neighbor and a $\big[ {+ \atop 1}
  \big]$ social/contact neighbor.  

We now illustrate how to build an approximate expression for each term of
Eq.~(\ref{Deltarho+1}) for $\Delta \rho^+\big|_{{- \atop 0} \to {+
    \atop 0}}$.  
For the sake of simplicity, we assume that all nodes have the same
number of neighbors $k=\mu$ chosen at random, 
which is equivalent to assuming that networks are degree-regular
random graphs.  However, we expect this approximation to work well
in networks with homogeneous degree distributions like ER networks.
The first term in Eq.~(\ref{Deltarho+1}) can be written as 
\begin{equation}
\Delta \rho^+\Big|_{{\overline{- \, +} \atop 0 ~ 0} \to {\overline{+ \, +}
    \atop 0 ~ 0} } = P\left({\mbox{\small $- \atop 0$}}\right) 
\sum_{\{\mathcal N^-_0 \}}^\mu M \left( \big\{\mathcal N^-_0 \big\},\mu \right)
  \frac{\mathcal N [\mbox{\tiny ${{{\overline{-+}} \atop
          0~0}}$}]}{\mu} \frac{1}{N},
\label{Deltarho+3}
\end{equation}
which can be understood as the product of the different probabilities
associated to each of the consecutive events that lead to the 
$\left[\mbox{\footnotesize ${{{\overline{-+}} \atop {0 \;0}}}$}\right]
\to \left[\mbox{\footnotesize ${{{\overline{++}} \atop {0
          \;0}}}$}\right]$ transition in a time step, as we describe
below.  A node $i$ 
with state $\big[ {- \atop 0} \big]$ is chosen at random with probability 
$P\left({\mbox{\small $- \atop 0$}}\right)$.  If node $i$ has 
$\mathcal N [\mbox{\tiny ${{{\overline{-+}} \atop 0~0}}$}]$ social
  neighbors in state $\big[ {+ \atop 0} \big]$, then one of these
  social neighbors $j$ is randomly chosen with probability 
$\mathcal N [\mbox{\tiny ${{{\overline{-+}} \atop 0~0}}$}]/\mu$, after
  which $i$ copies $j$' opinion with probability $1.0$ because
  there is no contact link between $i$ and $j$.  Finally, $\rho^+$
  increases by $1/N$ when $i$ switches opinion.  

\begin{figure}[t]
\begin{center}
  \includegraphics[width=7.5cm]{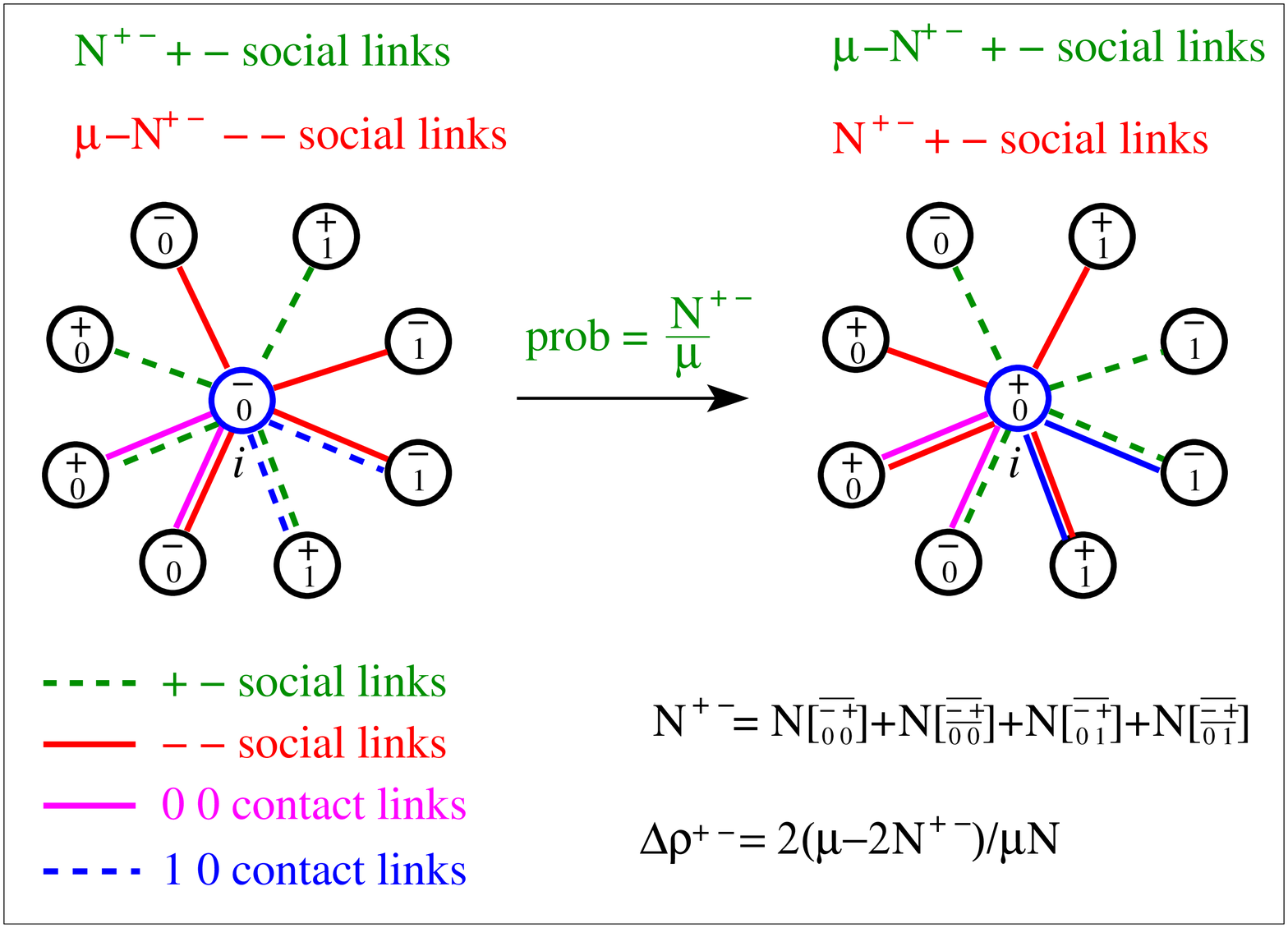}
  \caption{(Color online) Schematic illustration of an opinion update in which
    a node $i$ in state $\big[ {- \atop 0} \big]$ changes to state
    $\big[ {+ \atop 0} \big]$ by copying the opinion $+$ of a randomly
    chosen neighbor (green dashed links).  The change in the density of $+-$
    links is denoted by $\Delta \rho^{+-}$.}
\label{voter-update}
\end{center}
\end{figure}

In order to consider all possible scenarios of having $\mathcal N [\mbox{\tiny
    ${{{\overline{-+}} \atop 0~0}}$}]=0,1,..,\mu$ social neighbors we
sum over all possible neighborhood configurations represented by $\big
\{\mathcal N^-_0 \big\} \equiv \big\{ \mathcal N [\mbox{\tiny ${{{\overline{-+}}
      \atop {0~0}}}$}],  \mathcal N [\mbox{\tiny ${{{\overline{-+}}
      \atop \overline{0~0}}}$}], \mathcal N [\mbox{\tiny
  ${{{\overline{--}} \atop {0~0}}}$}],  \mathcal N [\mbox{\tiny
  ${{{\overline{--}} \atop \overline{0~0}}}$}],     \mathcal N
[\mbox{\tiny ${{{\overline{-+}} \atop {0~1}}}$}],  \mathcal N
[\mbox{\tiny ${{{\overline{-+}} \atop \overline{0~1}}}$}],     \mathcal
N [\mbox{\tiny ${{{\overline{--}} \atop {0~1}}}$}],  \mathcal N
[\mbox{\tiny ${{{\overline{--}} \atop \overline{0~1}}}$}] \big\}$,
weighted by the probability of each configuration 
$M \left( \big\{\mathcal N^-_0 \big\},\mu \right)$.  Here we denote by
$\mathcal N [\mbox{\tiny ${{{\overline{- \, 
          \mathcal O}} \atop {0 ~  \mathcal D}}}$}]$ the number of
$\big[ {\mathcal O \atop \mathcal D} \big]$ social neighbors and by
$\mathcal N [\mbox{\tiny ${{{\overline{- \, \mathcal O}} \atop
      \overline{0 ~  \mathcal D}}}$}]$ the number of $\big[ {\mathcal O
    \atop \mathcal D} \big]$ social/contact neighbors [see
  Fig.~(\ref{voter-update})].  The number of each type of neighbor is
between $0$ and 
$\mu$, and thus the sum in Eq.~(\ref{Deltarho+3}) include eight summations  
\begin{eqnarray*}
\sum_{\{\mathcal N^-_0 \}}^\mu \equiv 
\sum_{{\mathcal{~O}=+,- \atop \mathcal D=0,1}} 
\left( \sum_{\mathcal N \left[ \mbox{\tiny ${{{\overline{- \mathcal O}}
          \atop {0 \, \mathcal D}}}$} \right]=0}^\mu + 
\sum_{\mathcal N \left[ \mbox{\tiny ${{{\overline{- \, \mathcal O}}
          \atop \overline{0 ~  \mathcal D}}}$}\right]=0}^\mu \right) 
\end{eqnarray*}
over all combinations subject to the constraint 
\begin{eqnarray*}
\sum_{{\mathcal{~O}=+,- \atop \mathcal D=0,1}} 
\left( \mathcal N [\mbox{\tiny ${{{\overline{- \mathcal O}} \atop 
{0 \, \mathcal D}}}$}] + 
\mathcal N [\mbox{\tiny ${{{\overline{- \mathcal O}} \atop 
\overline{0 \, \mathcal D}}}$}] \right) = \mu.
\end{eqnarray*}

In order to carry out the summation in Eq.~(\ref{Deltarho+3}) we
only take into account correlations between first neighbors, and
neglect second and higher neighbor correlations (pair approximation).  Thus, we
define the probability 
$P[\mbox{\tiny ${{\overline{- \mathcal O} \atop {0 \, \mathcal
          D}}}$}] \equiv P \mbox{$\left({\rule[0.7mm]{3mm}{0.2mm} ~
    \mathcal O \atop  ~~~ 
    \mathcal D} \mbox{\large $|$} {- \atop 0}\right)$}$  
that a given
neighbor of node $i$ is a social neighbor with state $\big[ {\mathcal O \atop
    \mathcal D} \big]$, and consider 
$P[\mbox{\tiny ${{\overline{- \mathcal O} \atop {0 \, \mathcal
          D}}}$}]$
to be conditioned to the state $\big[ {- \atop 0} \big]$ of $i$ only,
and not on the other neighbors of $i$. Similarly, we denote by 
$P[\mbox{\tiny ${{\overline {- \mathcal O} \atop \overline{0 \,
          \mathcal D}}}$}] =  
P \mbox{$\left({\rule[0.7mm]{3mm}{0.2mm} ~ \mathcal O \atop
    \rule[0.7mm]{3mm}{0.2mm} ~ \mathcal D} \mbox{\large $|$} {- \atop
    0}\right)$}$ the
conditional probability that a node connected 
to $i$ is a social/contact neighbor with state $\big[ {\mathcal O \atop 
    \mathcal D} \big]$, given that $i$ has state $\big[ {- \atop 0}
  \big]$.  Therefore, $M$
becomes the multinomial probability distribution defined as 
\begin{eqnarray*}
M \left( \big\{\mathcal N^-_0 \big\},\mu \right) \equiv 
\begin{cases}
\frac{\mbox{\small $\mu!$}}{ { \mbox{\small $\prod$} \atop {\mbox{\tiny $\mathcal{~O}=+,- \atop \mathcal D=0,1$}}} \mbox{
    \footnotesize $\mathcal N$} [\mbox{\tiny 
    ${{{\overline{\mathcal O}} 
        \atop {\mathcal D}}}$}]!  
\mbox{ \footnotesize $\mathcal N$} [\mbox{\tiny ${{{\overline{
          \mathcal O}} \atop  
\overline{\mathcal D}}}$}]!} ~ {\mbox{\small $\prod$} \atop
  {\mbox{\tiny $\mathcal{~O}=+,- \atop \mathcal D=0,1$}}} 
P [\mbox{\tiny 
    ${{\overline {\mathcal O} 
        \atop {\mathcal D}}}$}]^{\mbox{ \footnotesize $\mathcal N$}
  [\mbox{\tiny  
    ${{{\overline{\mathcal O}} 
        \atop {\mathcal D}}}$}]} ~
P [\mbox{\tiny 
    ${{\overline {\mathcal O} 
        \atop \overline {\mathcal D}}}$}]^{\mbox{ \footnotesize $\mathcal N$}
  [\mbox{\tiny ${{{\overline{\mathcal O}} 
        \atop \overline {\mathcal D}}}$}]}  
& \mbox{when ${\mbox{\small $\sum$} \atop
  {\mbox{\tiny{$\mathcal {~O}=+,- \atop \mathcal D=0,1$}}}}
\left( \mathcal N [\mbox{\tiny ${{{\overline{ \mathcal O}} \atop
      {\mathcal D}}}$}] +   
\mathcal N [\mbox{\tiny ${{{\overline{\mathcal O}} \atop 
\overline{\mathcal D}}}$}] \right) = \mu$;}\\
0 & \text{otherwise,}
\end{cases}
\end{eqnarray*}
where we have used the symbols  
$\left[ \mbox{\footnotesize${\overline{\mathcal O} \atop \mathcal D}$}
    \right]$ and  
$\left[ \mbox{\footnotesize${\overline{\mathcal O} \atop
      \overline{\mathcal D}}$} \right]$ as 
short notations for  
$\big[\mbox{\footnotesize ${{{\overline{- \, \mathcal O}}
\atop {0 \;\mathcal D}}}$}\big]$ and
$\big[\mbox{\footnotesize ${{{\overline{- \, \mathcal O}}
\atop\overline{0 \;\mathcal D}}}$}\big]$, respectively.  Then,
performing the summation in Eq.~(\ref{Deltarho+3}) we arrive to 
\begin{equation}
\Delta \rho^+\Big|_{{\overline{- \, +} \atop 0 ~ 0} \to {\overline{+ \, +}
    \atop 0 ~ 0} }  =\frac{P\left({\mbox{\small $- \atop
      0$}}\right)  \langle \mathcal N [\mbox{\tiny
    ${{{\overline{-+}} \atop 0~0}}$}] \rangle}{\mu N} = 
\frac{P\left({\mbox{\small $- \atop 0$}}\right)  P [\mbox{\tiny
      ${{{\overline{-+}} \atop 0~0}}$}]}{N},  
\label{Deltarho+4}
\end{equation}
where we have used the identity $\langle \mathcal N [\mbox{\tiny
    ${{{\overline{-+}} \atop 0~0}}$}] \rangle = \mu \, P[\mbox{\tiny
    ${{\overline {- +} \atop {0 \; 0}}}$}]$ for the mean value of
$\mathcal N [\mbox{\tiny ${{{\overline{-+}} \atop 0~0}}$}]$.  
The other three terms in Eq.~(\ref{Deltarho+1}) can be obtained following an
approach similar to the one above for 
$\Delta \rho^+\Big|_{{\overline{- \, +} \atop 0 ~ 0} \to {\overline{+
      \, +}  \atop 0 ~ 0}}$, leading to the expression 
\begin{equation}
\Delta \rho^+\big|_{{- \atop 0} \to {+ \atop 0}} =    
\frac{P\left({\mbox{\small $- \atop 0$}}\right)}{N} \left( 
P [\mbox{\tiny ${{{\overline{-+}} \atop 0\;0}}$}] +
P [\mbox{\tiny ${{{\overline{-+}} \atop
        \overline{0 \; 0}}}$}] + 
P [\mbox{\tiny ${{{\overline{-+}} \atop 0 \;1}}$}] +
p_o \, P [\mbox{\tiny ${{{\overline{-+}} \atop
        \overline{0 \;1}}}$}] \right),
\label{Deltarho+5}
\end{equation}
where the prefactor $p_o$ in the last term accounts for the
probability of copying the opinion of an infected contact neighbor.
Keeping in mind that we aim to obtain a closed system of rate equations
for $\rho^+$, $\rho_1$, $\rho^{+-}$ and $\rho_{10}$, we now find
approximate expressions for the different probabilities of
Eq.~(\ref{Deltarho+5}) in 
terms of the fractions of nodes and links in each layer.  We start by
assuming that correlations between opinion and disease states of a
given node are negligible, and thus we can write   
\begin{equation}
P\left({\mbox{\small $- \atop 0$}}\right) \simeq \rho^- \rho_0.
\label{P-0}
\end{equation}
Then, to estimate the conditional probabilities $P[\mbox{\tiny ${{\overline 
        {- +} \atop {0 \, \mathcal  D}}}$}]$ and $P[\mbox{\tiny
    ${{\overline {- +} \atop \overline {0 \, \mathcal
          D}}}$}]$ it proves
convenient to split each of them into two conditional probabilities
\begin{eqnarray*}
P[\mbox{\tiny ${{\overline {- +} \atop {0 \, \mathcal D}}}$}]&=&
P \mbox{$\left({\rule[0.7mm]{3mm}{0.2mm} ~ + \atop  ~~~
    \mathcal D} \mbox{\large $|$} {- \atop 0}\right)$}=  
P \mbox{$\left({\rule[0.7mm]{3mm}{0.2mm} \atop } \mbox{\large
    $|$} {- \atop 0}\right)$}   
P \mbox{$\left( {+ \atop \mathcal D} \mbox{\large
    $|$} {\rule[0.7mm]{3mm}{0.2mm} \atop } {- \atop 0}\right)$}, \\
P[\mbox{\tiny ${{\overline {- +} \atop \overline{0 \, \mathcal D}}}$}]&=&
P \mbox{$\left({\rule[0.7mm]{3mm}{0.2mm} ~ + \atop
    \rule[0.7mm]{3mm}{0.2mm} ~ \mathcal D} \mbox{\large $|$} {- \atop
    0}\right)$}=   
P \mbox{$\left({\rule[0.7mm]{3mm}{0.2mm} \atop
    \rule[0.7mm]{3mm}{0.2mm} } \mbox{\large
    $|$} {- \atop 0}\right)$}   
P \mbox{$\left( {+ \atop \mathcal D} \mbox{\large
    $|$} {\rule[0.7mm]{3mm}{0.2mm} \atop \rule[0.7mm]{3mm}{0.2mm} } {-
    \atop 0}\right)$}, 
\end{eqnarray*}
using the relation $P(a,b|c)=P(a|c)P(b|a,c)$ and interpreting the
entire event of connecting a given type of link to a $[{+ \atop
  \mathcal D}]$ node as
two separate events.  Assuming that the type of link connected to node $i$ is 
uncorrelated with the state of $i$, we have  
\begin{eqnarray*}
P \mbox{$\left({\rule[0.7mm]{3mm}{0.2mm} \atop } \mbox{\large $|$} {- \atop
  0}\right)$} &\simeq& P \mbox{$\left({\rule[0.7mm]{3mm}{0.2mm}
  \atop}\right)$}=1-q ~~~~\mbox{and} \\ 
P \mbox{$\left({\rule[0.7mm]{3mm}{0.2mm} \atop
    \rule[0.7mm]{3mm}{0.2mm} } \mbox{\large $|$} {- \atop 0}\right)$}
&\simeq& P \mbox{$\left({\rule[0.7mm]{3mm}{0.2mm} \atop
    \rule[0.7mm]{3mm}{0.2mm} } \right)$}=q,
\end{eqnarray*}
and that opinion and disease states are uncorrelated, we have  
\begin{eqnarray*}
P \mbox{$ \left( {+ \atop \mathcal D} \mbox{\large
    $|$} {\rule[0.7mm]{3mm}{0.2mm} \atop } {- \atop 0}\right)$} &\simeq& 
P \left( {+ | \rule[1mm]{4mm}{0.2mm} \, -}\right) 
P \left( {\mathcal D | ~~~\, 0}\right) ~~~ \mbox{and} \\
P \mbox{$ \left( {+ \atop \mathcal D} \mbox{\large
    $|$} {\rule[0.7mm]{3mm}{0.2mm} \atop \rule[0.7mm]{3mm}{0.2mm} } 
{- \atop 0}\right)$} &\simeq& P \left( {+ | \rule[1mm]{4mm}{0.2mm} \, -}\right) 
P \left( {\mathcal D | \rule[1mm]{4mm}{0.2mm} \, 0}\right).
\end{eqnarray*}
Within an homogeneous pair approximation \cite{Vazquez-2008-2}, the
probability $P \left( {+ | \rule[1mm]{4mm}{0.2mm} \, -}\right)$ 
that a social neighbor $j$ of a node $i$ with opinion $\mathcal O_i=-$
has opinion 
$\mathcal O_j=+$ can be estimated as the ratio between the total number $\mu N 
\rho^{+-}/2$ of links from $-$ to $+$ nodes and the total number $\mu
N \rho^-$ of links connected to $-$ nodes, that is 
$P \left( {+ | \rule[1mm]{4mm}{0.2mm} \, -}\right) \simeq
\rho^{+-}/2\rho^-$.  Similarly, we estimate the
probability that a contact neighbor $j$ of a susceptible node has disease state
$\mathcal D_j=0$ as $P \left( {0 | \rule[1mm]{4mm}{0.2mm} \, 0}\right) \simeq
\rho_{00}/\rho_0$, and disease state $\mathcal D_j=1$ as 
$P \left( {1 | \rule[1mm]{4mm}{0.2mm} \, 0}\right) \simeq \rho_{10}/2\rho_0$.
And if $j$ is not a neighbor of $i$ on the contact layer then 
$P \left( {\mathcal D | ~~~\, 0}\right) \simeq \rho_{\mathcal D}$.  Assembling
all these factors we obtain
\begin{eqnarray}
P[\mbox{\tiny ${{\overline {- +} \atop {0 \; 0}}}$}] &\simeq& 
\frac{(1-q) \, \rho^{+-} \, \rho_0}{2\rho^-}, ~~~ 
P[\mbox{\tiny ${{\overline {- +} \atop \overline {0 \; 0}}}$}] \simeq 
\frac{q \, \rho^{+-} \, \rho_{00}}{2\rho^- \rho_0}, \nonumber \\ 
P[\mbox{\tiny ${{\overline {- +} \atop {0 \; 1}}}$}] &\simeq& 
\frac{(1-q) \, \rho^{+-} \, \rho_1}{2\rho^-}, ~~~
P[\mbox{\tiny ${{\overline {- +} \atop \overline {0 \; 1}}}$}] \simeq 
\frac{q \, \rho^{+-} \, \rho_{10}}{4\rho^- \rho_0}. 
\label{cond-prob1}
\end{eqnarray}
Finally, plugging into Eq.~(\ref{Deltarho+5}) the approximate expressions for
the conditional probabilities from Eqs.~(\ref{cond-prob1}) and for 
$P\left({\mbox{\small $- \atop 0$}}\right)$ from Eq.~(\ref{P-0}) we arrive to
\begin{equation} 
\Delta \rho^+\big|_{{- \atop 0} \to {+ \atop 0}} =
\frac{\rho^{+-}}{2N} \left[ \rho_0 - q(1-p_o) \frac{\rho_{10}}{2} \right],  
\label{Deltarho+-0+0}
\end{equation}
where we have used the conservation relations Eqs.~(\ref{c21}) and
(\ref{c24}).  

We now calculate the second gain term in Eq.~(\ref{drho+dt}), 
$\Delta \rho^+\big|_{{- \atop 1} \to {+ \atop 1}}$, which represents the
average change in $\rho^+$ due to 
$\big[ {- \atop 1} \big] \to \big[ {+ \atop 1} \big]$ transitions,
following the same steps as above for the term $\Delta \rho^+\big|_{{-
    \atop 0} \to {+ \atop 0}}$.  From Eq.~(\ref{Deltarho+2}) we obtain  
\begin{eqnarray*} 
\Delta \rho^+\big|_{{- \atop 1} \to {+ \atop 1}} &=&    
P\left({\mbox{\small $- \atop 1$}}\right) 
\sum_{\{\mathcal N^-_1 \}}^\mu M \left( \big\{ \mathcal N^-_1 \big\},\mu \right)
\frac{1}{\mu} \Big\{ \mathcal N [\mbox{\tiny ${{{\overline{-+}} \atop 1~0}}$}] + 
       p_o \, \mathcal N [\mbox{\tiny ${{{\overline{-+}} \atop \overline{1~0}}}$}] + 
       \mathcal N [\mbox{\tiny ${{{\overline{-+}} \atop 1~1}}$}] + 
       p_o \, \mathcal N [\mbox{\tiny ${{{\overline{-+}} \atop
          \overline{1~1}}}$}] \Big\} \frac{1}{N} \\
&=& \frac{P\left({\mbox{\small $- \atop 1$}}\right)}{\mu N} \Big\{
\langle \mathcal N [\mbox{\tiny ${{{\overline{-+}} \atop 1~0}}$}] \rangle +
\langle \mathcal N [\mbox{\tiny ${{{\overline{-+}} \atop 1~1}}$}] \rangle +
p_o \big( \langle \mathcal N [\mbox{\tiny ${{{\overline{-+}} \atop
          \overline{1~0}}}$}] \rangle + \langle \mathcal N [\mbox{\tiny
    ${{{\overline{-+}} \atop \overline{1~1}}}$}] \rangle \big) \Big\} \\
&=& \frac{P\left({\mbox{\small $- \atop 1$}}\right)}{N} \Big\{
P [\mbox{\tiny ${{{\overline{-+}} \atop 1~0}}$}] +
P [\mbox{\tiny ${{{\overline{-+}} \atop 1~1}}$}] +
p_o \big( P [\mbox{\tiny ${{{\overline{-+}} \atop
          \overline{1~0}}}$}] + P [\mbox{\tiny
    ${{{\overline{-+}} \atop \overline{1~1}}}$}] \big) \Big\},
\label{Deltarho+7}
\end{eqnarray*}
and using the approximations
\begin{eqnarray}
P[\mbox{\tiny ${{\overline {- +} \atop {1 \; 0}}}$}] &\simeq& 
\frac{(1-q) \, \rho^{+-} \, \rho_0}{2\rho^-}, ~~~
P[\mbox{\tiny ${{\overline {- +} \atop \overline {1 \; 0}}}$}] \simeq 
\frac{q \, \rho^{+-} \, \rho_{10}}{4 \rho^-  \rho_1}, \nonumber \\ 
P[\mbox{\tiny ${{\overline {- +} \atop {1 \; 1}}}$}] &\simeq& 
\frac{(1-q) \, \rho^{+-} \, \rho_1}{2\rho^-}, ~~~
P[\mbox{\tiny ${{\overline {- +} \atop \overline {1 \; 1}}}$}] \simeq 
\frac{q \, \rho^{+-} \, \rho_{11}}{2 \rho^- \rho_1},
\label{cond-prob2}
\end{eqnarray}
for the conditional probabilities we arrive to
\begin{equation} 
\Delta \rho^+\big|_{{- \atop 1} \to {+ \atop 1}} =
\frac{\rho^{+-} \rho_1}{2N} \left[ 1 - q(1-p_o) \right],
\label{Deltarho+-1+1}
\end{equation}
where we have used the conservation relations Eqs.~(\ref{c21}) and (\ref{c23}).

Adding Eqs.~(\ref{Deltarho+-0+0}) and (\ref{Deltarho+-1+1}) we obtain
the following expression for the average gain of a $+$ node in single
time step, corresponding to the sum of the first and second terms of
Eq.~(\ref{drho+dt}) 
\begin{equation}
\frac{d \rho^+}{dt}\Big|_{- \to +} =
\frac{1}{1/N} \left[ \Delta \rho^+\big|_{{- \atop 0} \to {+ \atop 0}} + 
\Delta \rho^+\big|_{{- \atop 1} \to {+ \atop 1}} \right] \simeq
\frac{1}{2}\omega \rho^{+-},
\label{Deltarho+gain}
\end{equation}
with
\begin{equation}
\omega \equiv 1-q(1-p_o) \left(\rho_1 + \frac{\rho_{10}}{2} \right).
\label{omega}
\end{equation}
The prefactor $\omega$ plays an important role in the dynamics of
opinion consensus, by setting the time scale 
associated to opinion updates, and can be interpreted as an effective
probability that a node adopts the opinion of randomly chosen opposite-opinion
neighbor.  That is,
Eq.~(\ref{Deltarho+gain}) for the gain of a $+$ 
node simply describes the process of selecting a $-$ node $i$ and a $+$
neighbor $j$, which happens with probability $\rho^{+-}/2$, and then 
switching $i$'s opinion with a probability $\omega$ that depends on
the connection type and disease state of both $i$ and $j$.
This ``effective copying probability'' $\omega$
turns out to be an average copying probability over the entire social
network, as shown in section \ref{analytical}.  

In order to find the equation for the average loss of a $+$ node in a
time step, corresponding to the sum of the third and forth terms of
Eq.~(\ref{drho+dt}), we can exploit the symmetry between $+$
and $-$ opinion states and simply interchange signs $+$ and $-$ in
Eq.~(\ref{Deltarho+gain})  
\begin{equation}
\frac{d \rho^+}{dt}\Big|_{+ \to -} =
\frac{1}{1/N} \left[ \Delta \rho^+\big|_{{+ \atop 0} \to {- \atop 0}} + 
\Delta \rho^+\big|_{{+ \atop 1} \to {- \atop 1}} \right] \simeq
-\frac{1}{2}\omega \rho^{+-},
\label{Deltarho+loss}
\end{equation}
where we used $\rho^{-+}=\rho^{+-}$.  Finally, adding
Eqs.~(\ref{Deltarho+gain}) and (\ref{Deltarho+loss}) we obtain 
\begin{equation}
\frac{d \rho^+}{dt}=0,
\end{equation}
quoted in Eq.~(\ref{Drho+Dt}) of the main text.  Therefore, the
fractions of $+$ and $-$ nodes are conserved at all times:
$\rho^+(t)=\rho^+(0)$ and $\rho^-(t)=\rho^-(0)=1-\rho^+(0)$.  Even
though the above calculation leads to a very simple result, it serves
as an introduction to the methodology used for deriving rate equations
for the other fractions $\rho^{+-}$, $\rho_{1}$ and $\rho_{10}$, as we show next.

\section{Derivation of the rate equation for $\rho^{+-}$}
\label{rho+-}

In analogy to the calculation for $\rho^+$ in the previous section,
the average change of the faction of $+-$ social links $\rho^{+-}$ in 
a time step is given by the rate equation
\begin{eqnarray}
\frac{d \rho^{+-}}{dt} = \frac{d \rho^{+-}}{dt}\Big|_{-\to+} + 
\frac{d \rho^{+-}}{dt}\Big|_{+\to-},
\label{drho+-dt}
\end{eqnarray}
with
\begin{eqnarray}
\label{drho+-dt-+}
\frac{d \rho^{+-}}{dt}\Big|_{-\to+} &=& \frac{1}{1/N} \left[ 
\Delta \rho^{+-}\big|_{{- \atop 0} \to {+ \atop 0}} +   
\Delta \rho^{+-}\big|_{{- \atop 1} \to {+ \atop 1}} \right] \\
\label{drho+-dt+-}
\frac{d \rho^{+-}}{dt}\Big|_{+\to-} &=& \frac{1}{1/N} \left[   
\Delta \rho^{+-}\big|_{{+ \atop 0} \to {- \atop 0}} +  
\Delta \rho^{+-}\big|_{{+ \atop 1} \to {- \atop 1}} \right]= 
\left\{ \frac{d \rho^{+-}}{dt}\Big|_{-\to+} \right\}^{- \Longleftrightarrow +},
\end{eqnarray}
where the symbol $- \Longleftrightarrow +$ indicates the interchange
of signs $+$ and $-$ in the expression between braces.
Equation~(\ref{drho+-dt+-}) means that the symmetry between $+$ and $-$
opinions allows to find the second term in Eq.~(\ref{drho+-dt}) by
interchanging signs in the first term.  To calculate the first term in
Eq.~(\ref{drho+-dt-+}) we sum over all four types of interactions 
of a $\big[{- \atop 0}\big]$ node $i$ with a $\big[{+ \atop \mathcal D}\big]$ 
neighbor $j$ that lead to the 
$\big[ {- \atop 0} \big] \to \big[ {+ \atop 0} \big]$ transition  
\begin{eqnarray}
\Delta \rho^{+-}\big|_{{- \atop 0} \to {+ \atop 0}} &=&    
\Delta \rho^{+-}\Big|_{{\overline{- +} \atop 0 \; 0} \to {\overline{+ +}
    \atop 0 \; 0} } +    
\Delta \rho^{+-}\Big|_{{\overline{- +} \atop  \overline{0 \; 0}} \to
  {\overline{+ +} \atop \overline{0 \; 0}}} +
\Delta \rho^{+-}\Big|_{{\overline{- +} \atop 0 \; 1} \to {\overline{+ +}
    \atop 0 \; 1} } +    
\Delta \rho^{+-}\Big|_{{\overline{- +} \atop  \overline{0 \; 1}} \to
  {\overline{+ +} \atop \overline{0 \; 1}}}. 
\end{eqnarray}
As explained in the previous section, the probabilities of
interactions $[\mbox{\scriptsize ${{{\overline{-+}} \atop
        {0 \,\mathcal D}}}$}]$ and $[\mbox{\scriptsize
    ${{{\overline{-+}} \atop \overline{0 \,\mathcal D}}}$}]$ are given
by the respective fractions $\mathcal N [\mbox{\tiny ${{{\overline{-+}} \atop
        {0\,\mathcal D}}}$}]/\mu$ and $\mathcal N
[\mbox{\tiny ${{{\overline{-+}} \atop \overline{0\,\mathcal
          D}}}$}]/\mu$ of each type of neighbor.
 The change in the number of $+-$ social links after
node $i$ switches opinion is given by the expression 
$\mu -2 \big( \mathcal N [\mbox{\tiny ${{{\overline{-+}} \atop 0~0}}$}] + 
\mathcal N [\mbox{\tiny ${{{\overline{-+}} \atop \overline{0~0}}}$}]+
\mathcal N [\mbox{\tiny ${{{\overline{-+}} \atop 0~1}}$}] + 
\mathcal N [\mbox{\tiny ${{{\overline{-+}} \atop \overline{0~1}}}$}]
\big)$, which takes into account the specific configuration of links
and neighbors connected to $i$, as depicted in Fig.~\ref{voter-update}.
We obtain     
 \begin{eqnarray}
\Delta \rho^{+-}\big|_{{- \atop 0} \to {+ \atop 0}} 
&=& P\left({\mbox{\small $- \atop 0$}}\right) 
\sum_{\{\mathcal N^-_0 \}}^\mu 
\frac{M \left( \big\{\mathcal N^-_0 \big\},\mu \right)}{\mu} \big( 
       \mathcal N [\mbox{\tiny ${{{\overline{-+}} \atop 0~0}}$}] + 
       \mathcal N [\mbox{\tiny ${{{\overline{-+}} \atop \overline{0~0}}}$}]+
       \mathcal N [\mbox{\tiny ${{{\overline{-+}} \atop 0~1}}$}] + 
p_o \, \mathcal N [\mbox{\tiny ${{{\overline{-+}} \atop \overline{0~1}}}$}]
\big) \nonumber \\ 
&\times&\frac{ \big[\mu -2 \big( \mathcal N [\mbox{\tiny ${{{\overline{-+}} \atop 0~0}}$}] + 
\mathcal N [\mbox{\tiny ${{{\overline{-+}} \atop \overline{0~0}}}$}]+
\mathcal N [\mbox{\tiny ${{{\overline{-+}} \atop 0~1}}$}] + 
\mathcal N [\mbox{\tiny ${{{\overline{-+}} \atop \overline{0~1}}}$}]
\big) \big]}{\mu N/2} \\ 
&=& \frac{2 P\left( {\mbox{\small $- \atop
      0$}}\right)}{\mu^2 N}  \bigg\{ \mu \Big[ 
\langle \mathcal N [\mbox{\tiny ${{{\overline{-+}} \atop 0~0}}$}] \rangle + 
\langle \mathcal N [\mbox{\tiny ${{{\overline{-+}} \atop
        \overline{0~0}}}$}] \rangle +
\langle \mathcal N [\mbox{\tiny ${{{\overline{-+}} \atop 0~1}}$}] \rangle + 
p_o \, \langle \mathcal N [\mbox{\tiny ${{{\overline{-+}} \atop
        \overline{0~1}}}$}] \rangle \Big] \nonumber \\
&-& 2 \Big[   
\big \langle \mathcal N [\mbox{\tiny ${{{\overline{-+}} \atop
        0~0}}$}]^2 \big \rangle +  
\big \langle \mathcal N [\mbox{\tiny ${{{\overline{-+}} \atop
        \overline{0~0}}}$}]^2 \big \rangle +
\big \langle \mathcal N [\mbox{\tiny ${{{\overline{-+}} \atop
        0~1}}$}]^2 \rangle +  
p_o \, \langle \mathcal N [\mbox{\tiny ${{{\overline{-+}} \atop
        \overline{0~1}}}$}]^2 \big \rangle \nonumber \\
&+& 2 \big(     
\langle \mathcal N [\mbox{\tiny ${{{\overline{-+}} \atop 0~0}}$}]
\mathcal N [\mbox{\tiny ${{{\overline{-+}} \atop \overline{0~0}}}$}] \rangle + 
\langle \mathcal N [\mbox{\tiny ${{{\overline{-+}} \atop 0~0}}$}]
\mathcal N [\mbox{\tiny ${{{\overline{-+}} \atop {0~1}}}$}] \rangle + 
\langle \mathcal N [\mbox{\tiny ${{{\overline{-+}} \atop \overline{0~0}}}$}]
\mathcal N [\mbox{\tiny ${{{\overline{-+}} \atop {0~1}}}$}] \rangle
\big) \nonumber \\
&+& (1+p_o) \big(   
\langle \mathcal N [\mbox{\tiny ${{{\overline{-+}} \atop 0~0}}$}]
\mathcal N [\mbox{\tiny ${{{\overline{-+}} \atop \overline{0~1}}}$}] \rangle + 
\langle \mathcal N [\mbox{\tiny ${{{\overline{-+}} \atop \overline{0~0}}}$}]
\mathcal N [\mbox{\tiny ${{{\overline{-+}} \atop \overline{0~1}}}$}] \rangle + 
\langle \mathcal N [\mbox{\tiny ${{{\overline{-+}} \atop {0~1}}}$}]
\mathcal N [\mbox{\tiny ${{{\overline{-+}} \atop \overline{0~1}}}$}] \rangle
\big) \Big] \bigg\},
\label{Deltarho+--0+0}
\end{eqnarray}
where the first and second moments of 
$M \left( \big\{\mathcal N^-_0 \big\},\mu \right)$ are
\begin{eqnarray}
\langle \mathcal N [\mbox{\tiny ${\overline{- ~ +} \atop \mathcal D_i \,
        \mathcal D_j}$}] \rangle &=& \mu 
P [\mbox{\tiny ${\overline{- ~ +} \atop \mathcal D_i \, \mathcal
        D_j}$}], \nonumber \\
\langle \mathcal N [\mbox{\tiny ${\overline{- ~ +} \atop
      \overline{\mathcal D_i \, \mathcal D_j}}$}] \rangle &=& \mu 
P [\mbox{\tiny ${\overline{- ~ +} \atop \overline{\mathcal D_i \, \mathcal
        D_j}}$}], \nonumber \\  
\langle \mathcal N [\mbox{\tiny ${\overline{- ~ +} \atop \mathcal D_i \,
        \mathcal D_j}$}]^2 \rangle &=& \mu 
P [\mbox{\tiny ${\overline{- ~ +} \atop \mathcal D_i \, \mathcal 
        D_j}$}] + \mu(\mu-1) P [\mbox{\tiny ${\overline{- ~ +}
        \atop \mathcal D_i \, \mathcal D_j}$}]^2, \nonumber \\
\langle \mathcal N [\mbox{\tiny ${\overline{- ~ +} \atop
      \overline{\mathcal D_i \, \mathcal D_j}}$}]^2 \rangle &=& \mu  
P [\mbox{\tiny ${\overline{- ~ +} \atop \overline{\mathcal D_i \, \mathcal 
        D_j}}$}] + \mu(\mu-1) P [\mbox{\tiny ${\overline{- ~ +}
        \atop \overline{\mathcal D_i \, \mathcal D_j}}$}]^2,\nonumber \\ 
\langle \mathcal N [\mbox{\tiny ${\overline{- ~ +} \atop \mathcal D_i \,
        \mathcal D_j}$}] \, \mathcal N [\mbox{\tiny ${\overline{- ~
            +} \atop \mathcal D_i \, \mathcal D_j'}$}] \rangle &=&  
        \mu(\mu-1) P [\mbox{\tiny ${\overline{- ~ +}
        \atop \mathcal D_i \, \mathcal D_j}$}] \, P [\mbox{\tiny
            ${\overline{- ~ +} \atop \mathcal D_i \, \mathcal
                D_j'}$}] , \nonumber \\
\langle \mathcal N [\mbox{\tiny ${\overline{- ~ +} \atop
        \overline{\mathcal D_i \, \mathcal D_j}}$}] \, \mathcal N
        [\mbox{\tiny ${\overline{- ~ +} \atop \overline{\mathcal D_i \,
              \mathcal D_j'}}$}] \rangle &=&   
        \mu(\mu-1) P [\mbox{\tiny ${\overline{- ~ +}
        \atop \overline{\mathcal D_i \, \mathcal D_j}}$}] \, P [\mbox{\tiny
            ${\overline{- ~ +} \atop \overline{\mathcal D_i \,
                \mathcal D_j' }}$}] , \nonumber \\ 
\langle \mathcal N [\mbox{\tiny ${\overline{- ~ +} \atop
        \mathcal D_i \, \mathcal D_j}$}] \, \mathcal N
        [\mbox{\tiny ${\overline{- ~ +} \atop \overline{\mathcal D_i \,
              \mathcal D_j'}}$}] \rangle &=&   
        \mu(\mu-1) P [\mbox{\tiny ${\overline{- ~ +}
        \atop {\mathcal D_i \, \mathcal D_j}}$}] \, P [\mbox{\tiny
            ${\overline{- ~ +} \atop \overline{\mathcal D_i \,
                \mathcal D_j'}}$}].
\label{moments}
\end{eqnarray}
Here $\mathcal D_i=1,0$ and $\mathcal D_j=1,0$ are the disease states
of nodes $i$ and $j$, respectively.  Replacing the expressions for the
moments from Eqs.~(\ref{moments}) in Eq.~(\ref{Deltarho+--0+0}) and
regrouping terms we obtain 
 \begin{eqnarray}
\Delta \rho^{+-}\big|_{{- \atop 0} \to {+ \atop 0}}= 
&=& \frac{2 P\left( {\mbox{\small $- \atop
      0$}}\right)}{\mu N}  \bigg\{ (\mu-2) \Big[ 
P[\mbox{\tiny ${\overline{-+} \atop 0~0}$}] + 
P [\mbox{\tiny ${\overline{-+} \atop
        \overline{0~0}}$}] +
P [\mbox{\tiny ${\overline{-+} \atop 0~1}$}] + 
p_o \, P [\mbox{\tiny ${\overline{-+} \atop
        \overline{0~1}}$}] \Big] \nonumber \\
&-& 2 (\mu-1) \Big[ \big(  
P [\mbox{\tiny ${\overline{-+} \atop 0~0}$}] +  
P [\mbox{\tiny ${\overline{-+} \atop \overline{0~0}}$}] +
P [\mbox{\tiny ${\overline{-+} \atop 0~1}$}] \big)^2 +
p_o \, P [\mbox{\tiny ${\overline{-+} \atop \overline{0~1}}$}]^2
\nonumber \\
&+& (1+p_o) P [\mbox{\tiny ${\overline{-+} \atop \overline{0~1}}$}]  \big(     
P [\mbox{\tiny ${\overline{-+} \atop 0~0}$}] + 
P [\mbox{\tiny ${\overline{-+} \atop \overline{0~0}}$}] + 
P [\mbox{\tiny ${\overline{-+} \atop 0~1}$}] \big) \Big] \bigg\}. 
\label{Deltarho+--0+02}
\end{eqnarray}
Plugging the expressions for the probabilities 
$P [\mbox{\tiny ${\overline{-+} \atop 0~\mathcal D}$}]$ and 
$P [\mbox{\tiny ${\overline{-+} \atop \overline{0~\mathcal D}}$}]$ from
  Eq.~(\ref{cond-prob1}) into Eq.~(\ref{Deltarho+--0+02}), and after
  doing some algebra we finally obtain 
\begin{eqnarray}
\Delta \rho^{+-}\big|_{{- \atop 0} \to {+ \atop 0}} =
\frac{\rho^{+-}}{\mu N} \bigg\{ (\mu-2) \left[ (1-q) \rho_0 + q \left(
  \rho_{00} + p_o \frac{\rho_{10}}{2} \right) \right] - 
(\mu-1) \frac{\rho^{+-}}{\rho^-} \left[\rho_0 -(1-p_o) q
  \frac{\rho_{10}}{2} \right] \bigg\}. \nonumber \\ 
\label{Deltarho+--0+03}
\end{eqnarray}
We now follow an approach similar to the one above for 
$\Delta \rho^{+-}\big|_{{- \atop 0} \to {+ \atop 0}}$ and calculate
the second term of Eq.~(\ref{drho+-dt-+}) as
 \begin{eqnarray}
\Delta \rho^{+-}\big|_{{- \atop 1} \to {+ \atop 1}} 
&=& P\left({\mbox{\small $- \atop 1$}}\right) 
\sum_{\{\mathcal N^-_1 \}}^\mu 
\frac{M \left( \big\{\mathcal N^-_1 \big\},\mu \right)}{\mu} \big[ 
       \mathcal N [\mbox{\tiny ${\overline{-+} \atop 1~0}$}] + 
       \mathcal N [\mbox{\tiny ${\overline{-+} \atop 1~1}$}] + 
p_o \big( \mathcal N [\mbox{\tiny ${\overline{-+} \atop \overline{1~0}}$}]+
        \mathcal N [\mbox{\tiny ${\overline{-+} \atop
              \overline{1~1}}$}] \big) \big] \nonumber \\ 
&\times&\frac{ \big[\mu -2 \big( \mathcal N [\mbox{\tiny
        ${\overline{-+} \atop 1~0}$}] +  
\mathcal N [\mbox{\tiny ${\overline{-+} \atop \overline{1~0}}$}]+
\mathcal N [\mbox{\tiny ${\overline{-+} \atop 1~1}$}] + 
\mathcal N [\mbox{\tiny ${\overline{-+} \atop \overline{1~1}}$}]
\big) \big]}{\mu N/2} \nonumber \\ 
&=& \frac{2 P\left( {\mbox{\small $- \atop
      1$}}\right)}{\mu^2 N}  \bigg\{ \mu \Big[ 
\langle \mathcal N [\mbox{\tiny ${\overline{-+} \atop 1~0}$}] \rangle + 
\langle \mathcal N [\mbox{\tiny ${\overline{-+} \atop 1~1}$}] \rangle + 
p_o \big( \langle \mathcal N [\mbox{\tiny ${\overline{-+} \atop
        \overline{1~0}}$}] \rangle +
\langle \mathcal N [\mbox{\tiny ${\overline{-+} \atop
        \overline{1~1}}$}] \rangle \big) \Big] \nonumber \\
&-& 2 \Big[   
\big \langle \mathcal N [\mbox{\tiny ${\overline{-+} \atop
        1~0}$}]^2 \big \rangle +  
\big \langle \mathcal N [\mbox{\tiny ${\overline{-+} \atop
        1~1}$}]^2 \rangle +  
p_o \big( \big \langle \mathcal N [\mbox{\tiny ${\overline{-+} \atop
        \overline{1~0}}$}]^2 \big \rangle +
\langle \mathcal N [\mbox{\tiny ${\overline{-+} \atop
        \overline{1~1}}$}]^2 \big \rangle \big) + 
2 \langle \mathcal N [\mbox{\tiny ${\overline{-+} \atop 1~0}$}]
\mathcal N [\mbox{\tiny ${\overline{-+} \atop 1~1}$}] \rangle \nonumber \\
&+& (1+p_o) \big( \langle \mathcal N [\mbox{\tiny ${\overline{-+} \atop 1~0}$}]
\mathcal N [\mbox{\tiny ${\overline{-+} \atop \overline{1~0}}$}] \rangle + 
\langle \mathcal N [\mbox{\tiny ${\overline{-+} \atop 1~0}$}]
\mathcal N [\mbox{\tiny ${\overline{-+} \atop \overline{1~1}}$}]
\rangle + 
\langle \mathcal N [\mbox{\tiny ${\overline{-+} \atop \overline{1~0}}$}]
\mathcal N [\mbox{\tiny ${\overline{-+} \atop 1~1}$}] \rangle + 
\langle \mathcal N [\mbox{\tiny ${\overline{-+} \atop 1~1}$}]
\mathcal N [\mbox{\tiny ${\overline{-+} \atop \overline{1~1}}$}]
\rangle \big) \nonumber \\
&+& 2 p_o \langle \mathcal N [\mbox{\tiny ${\overline{-+} \atop
      \overline{1~0}}$}] 
\mathcal N [\mbox{\tiny ${\overline{-+} \atop \overline{1~1}}$}] \rangle
\big) \Big] \bigg\} \nonumber \\
&=& \frac{2 P\left( {\mbox{\small $- \atop
      1$}}\right)}{\mu N}  \bigg\{ (\mu-2) \Big[ 
P[\mbox{\tiny ${\overline{-+} \atop 1~0}$}] + 
P [\mbox{\tiny ${\overline{-+} \atop 1~1}$}] + 
p_o \big( P [\mbox{\tiny ${\overline{-+} \atop \overline{1~0}}$}] + 
P [\mbox{\tiny ${\overline{-+} \atop \overline{1~1}}$}] \big) \Big] 
\nonumber \\
&-& 2 (\mu-1) \Big[ \big(  
P [\mbox{\tiny ${\overline{-+} \atop 1~0}$}] +
P [\mbox{\tiny ${\overline{-+} \atop 1~1}$}] \big)^2 +
p_o \big( P [\mbox{\tiny ${\overline{-+} \atop \overline{1~0}}$}] +
P [\mbox{\tiny ${\overline{-+} \atop \overline{1~1}}$}] \big)^2 +
\nonumber \\
&+& (1+p_o) \big( 
P [\mbox{\tiny ${\overline{-+} \atop 1~0}$}] +       
P [\mbox{\tiny ${\overline{-+} \atop 1~1}$}] \big) \big( 
P [\mbox{\tiny ${\overline{-+} \atop \overline{1~0}}$}] + 
P [\mbox{\tiny ${\overline{-+} \atop 1~1}$}] \big) \Big] \bigg\}, 
\label{Deltarho+--1+1}
\end{eqnarray}
where we have used the moments from Eqs.~(\ref{moments}).  After
substituting expressions (\ref{cond-prob2}) for the probabilities 
$P [\mbox{\tiny ${\overline{-+} \atop 1~\mathcal D}$}]$ and 
$P [\mbox{\tiny ${\overline{-+} \atop \overline{1~\mathcal D}}$}]$ 
we arrive to
\begin{eqnarray}
\Delta \rho^{+-}\big|_{{- \atop 1} \to {+ \atop 1}} =
\frac{\rho^{+-}}{\mu N} \bigg\{ (\mu-2) \left[ (1-q) \rho_1 + q \, p_o \left(
\rho_{11} + \frac{\rho_{10}}{2} \right) \right] - 
(\mu-1) \frac{\rho^{+-} \rho_1}{\rho^-} \left[1- (1-p_o) q \right]
\bigg\}. \nonumber \\
\label{Deltarho+--1+12}
\end{eqnarray}
By adding Eqs.~(\ref{Deltarho+--0+03}) and (\ref{Deltarho+--1+12}) we
obtain the following expression for the change in $\rho^{+-}$ due to 
$- \to +$ transitions
\begin{eqnarray}
\frac{d \rho^{+-}}{dt}\Big|_{-\to+} = \frac{\omega \rho^{+-}}{\mu} 
\left[ \mu - 2 - (\mu-1) \frac{\rho^{+-}}{\rho^-} \right].
\label{drho+-dt-+1}
\end{eqnarray}
Then, by interchanging sings $+$ and $-$ in Eq.~(\ref{drho+-dt-+1}) we
obtain the change in $\rho^+$ due to $+ \to -$ transitions 
\begin{eqnarray}
\frac{d \rho^{+-}}{dt}\Big|_{+\to-} = \frac{\omega \rho^{+-}}{\mu} 
\left[ \mu - 2 - (\mu-1) \frac{\rho^{+-}}{\rho^+} \right].
\label{drho+-dt+-1}
\end{eqnarray}
Finally, adding Eqs.~(\ref{drho+-dt-+1}) and (\ref{drho+-dt+-1}) we
arrive to the following rate equation for $\rho^{+-}$ quoted in
Eq.~(\ref{Drho+-Dt}) of the main text 
\begin{eqnarray*}
\frac{d\rho^{+-}}{dt}&=&\frac{2 \omega \rho^{+-}}{\mu} \left[ (\mu-1) 
\left(1-\frac{\rho^{+-}}{2 \rho^+ \rho^-} \right) -1 \right].
\end{eqnarray*}

\section{Derivation of the rate equation for $\rho_1$}
\label{rho1}

The average change of the fraction of infected nodes $\rho_1$ in a
single time step can be written as 
\begin{eqnarray}
\frac{d \rho_1}{dt} = \frac{1}{1/N} \left[ 
\Delta \rho_1\big|_{{+ \atop 1} \to {+ \atop 0}} +  
\Delta \rho_1\big|_{{+ \atop 0} \to {+ \atop 1}} +   
\Delta \rho_1\big|_{{- \atop 1} \to {- \atop 0}} + 
\Delta \rho_1\big|_{{- \atop 0} \to {- \atop 1}} \right], 
\label{drho1dt}
\end{eqnarray}
where each term represents a different transition corresponding to a
disease update on the contact layer.  The first term of
Eq.~(\ref{drho1dt}) corresponds to the recovery of a 
$\big[{+ \atop 1}\big]$ node, and can be estimated as 
\begin{equation}
\Delta \rho_1\big|_{{+ \atop 1} \to {+ \atop 0}} = -  
P\left({\mbox{\small $+ \atop 1$}}\right) (1-\beta) 
\frac{1}{N} \simeq -\frac{(1-\beta)}{N} \rho^+ \rho_1.
\label{Deltarho1+1}
\end{equation}
That is, with probability $P\left({\mbox{\small $+ \atop 1$}}\right)
\simeq \rho^+ \rho_1$ a $\big[{+ \atop 1}\big]$ node is picked at random,
and then recovers with probability $1-\beta$, decreasing $\rho_1$
in $1/N$.  The second term corresponds to the infection of a 
$\big[{+ \atop 0}\big]$ node, while the last two terms are
equivalent to the first two, but where a node with opinion $-$ is
recovered and infected, respectively.  By the symmetry of $+$ and $-$
opinions, the last two terms are obtained by interchanging signs $+$ and $-$
in the first two.  

We now find an approximate expression for the second term of
Eq.~(\ref{drho1dt}).  A susceptible node $j$ in state  
$\big[{+ \atop 0}\big]$ can be
infected by a sick neighbor $i$ with $+$ or $-$ opinion and connected
to $j$ by a contact link or by both a social and a contact link.  Thus,
four possible contact interactions lead to the 
$\big[{+ \atop 0}\big] \to \big[{+ \atop 1}\big]$ transition:
\begin{equation}
\Delta \rho_1\big|_{{+ \atop 0} \to {+ \atop 1}} =    
\Delta \rho_1\Big|_{{+ + \atop \overline{1 \; 0}} \to {+ +
    \atop \overline{1 \; 1}}} +    
\Delta \rho_1\Big|_{{\overline{+ +} \atop  \overline{1 \; 0}} \to
  {\overline{+ +} \atop \overline{1 \; 1}}} +
\Delta \rho_1\Big|_{{- + \atop \overline{1 \; 0}} \to {- +
    \atop \overline{1 \; 1}}} +    
\Delta \rho_1\Big|_{{\overline{- +} \atop  \overline{1 \; 0}} \to
  {\overline{- +} \atop \overline{1 \; 1}}}.
\label{Deltarho1+0}
\end{equation}
The symbol $\big[\mbox{\footnotesize ${{\overline{\mathcal O \, +}}
\atop {1 \; 0}}$}\big]$ represents a contact interaction between node
$i$ in state   
$\big[{\mathcal O \atop 1}\big]$ ($\mathcal O=+,-$) and node $j$
in state $\big[{+ \atop 0}\big]$.  The state that changes in the
interaction is now displayed on the right-hand side 
of the symbol, instead on the left-hand side as for the case of the social
interactions described in the previous sections.  This is because the
chosen neighbor $j$ of $i$ changes state in the CP, while in the VM is
node $i$ who changes state.  Taking into account the events 
and their associated probabilities that lead to each of the four
interactions described above, we can write Eq.~(\ref{Deltarho1+0}) as 
\begin{eqnarray}
\Delta \rho_1\big|_{{+ \atop 0} \to {+ \atop 1}} &=& 
P\left({\mbox{\small $+ \atop 1$}}\right) 
\sum_{\{\mathcal N^+_1 \}}^\mu M \left( \big\{\mathcal N^+_1
  \big\},\mu \right) \frac{\beta}{\mu} \big( 
\mathcal N [\mbox{\tiny ${+ + \atop \overline{1~0}}$}] +  
\mathcal N [\mbox{\tiny ${\overline{+ +} \atop \overline{1~0}}$}]
\big) \frac{1}{N} \nonumber \\ &+&
P\left({\mbox{\small $- \atop 1$}}\right) 
\sum_{\{\mathcal N^-_1 \}}^\mu M \left( \big\{\mathcal N^-_1
  \big\},\mu \right) \frac{\beta}{\mu} \big( 
\mathcal N [\mbox{\tiny ${- + \atop \overline{1~0}}$}] +  
p_d \, \mathcal N [\mbox{\tiny ${\overline{- +} \atop \overline{1~0}}$}]
\big) \frac{1}{N}.
\label{Deltarho1+01} 
\end{eqnarray}
The first and third terms of Eq.~(\ref{Deltarho1+01}) correspond to selecting an
$\big[{\mathcal O \atop 1}\big]$ node $i$  and a contact neighbor $j$
with state $\big[{+ \atop 0}\big]$ at random, which happens with probability
$P\left({\mbox{\small $\mathcal O \atop 1$}}\right)  
\mathcal N [\mbox{\tiny ${\mathcal O + \atop \overline{1~0}}$}]/\mu$,
and then $i$ infecting $j$ with probability $\beta$, given that they
are not connected by a social link.  The second and fourth terms are
similar to the first and second terms, respectively, but selecting a
social/contact neighbor $j$.  As both types of links are present in
this case, $i$ infects $j$ with probability $\beta \, p_d$ when both nodes have
different opinions (fourth term).  In all cases $\rho_1$ changes by
$1/N$.  Performing the sums of Eq.~(\ref{Deltarho1+01}) we obtain 
\begin{eqnarray}
\Delta \rho_1\big|_{{+ \atop 0} \to {+ \atop 1}} = 
\frac{\beta}{\mu N} \Big[ 
P\left({\mbox{\small $+ \atop 1$}}\right) \big( 
\langle \mathcal N [\mbox{\tiny ${+ + \atop \overline{1~0}}$}]\rangle +  
\langle \mathcal N [\mbox{\tiny ${\overline{+ +} \atop
      \overline{1~0}}$}] \rangle \big) + 
P\left({\mbox{\small $- \atop 1$}}\right) \big(
\langle \mathcal N [\mbox{\tiny ${- + \atop \overline{1~0}}$}] \rangle+  
p_d \, \langle \mathcal N [\mbox{\tiny ${\overline{- +} \atop
      \overline{1~0}}$}] \rangle \big) \Big]. 
\label{Deltarho1+02} 
\end{eqnarray}
Replacing the expressions for the first moments 
$\langle \mathcal N [\mbox{\tiny ${\mathcal O + \atop
\overline{1~0}}$}]\rangle= \mu P[\mbox{\tiny ${\mathcal O + \atop
      \overline{1~0}}$}]$ and 
$\langle \mathcal N [\mbox{\tiny ${\overline{\mathcal O +} \atop
\overline{1~0}}$}]\rangle= \mu P[\mbox{\tiny ${\overline{\mathcal O +}\atop
      \overline{1~0}}$}]$ in Eq.~(\ref{Deltarho1+02}), and using the
following expressions for the conditional probabilities 
\begin{eqnarray}
\label{cond-prob3}
P[\mbox{\tiny ${+ + \atop \overline{1 \; 0}}$}] &\simeq& 
\frac{(1-q) \, \rho^+ \, \rho_{10}}{2\rho_1}, ~~~
P[\mbox{\tiny ${\overline{+ +} \atop \overline {1 \; 0}}$}] \simeq 
\frac{q \, \rho^{++} \, \rho_{10}}{2\rho^+ \rho_1}, \\ 
\label{cond-prob4}
P[\mbox{\tiny ${- + \atop \overline{1 \; 0}}$}] &\simeq& 
\frac{(1-q) \, \rho^+ \, \rho_{10}}{2\rho_1}, ~~~
P[\mbox{\tiny ${\overline{- +} \atop \overline {1 \; 0}}$}] \simeq 
\frac{q \, \rho^{+-} \, \rho_{10}}{4\rho^- \rho_1}, 
\end{eqnarray}
we finally arrive to  
\begin{equation}
\Delta \rho_1\big|_{{+ \atop 0} \to {+ \atop 1}} \simeq \frac{\beta \,
  \rho_{10}}{2N} \left[ \rho^+ - \frac{q}{2}(1-p_d) \rho^{+-} \right], 
\label{Deltarho1+03} 
\end{equation}
where we have used the conservation relations from Eqs.~(\ref{c11}) and
(\ref{c13}).  

Now that we estimated the first two terms of Eq.~(\ref{drho1dt}), the
last two terms are obtained by interchanging signs $+$ and $-$ in
Eqs.~(\ref{Deltarho1+1}) and (\ref{Deltarho1+03}):  
\begin{eqnarray}
\label{Deltarho1-1} 
\Delta \rho_1\big|_{{- \atop 1} \to {- \atop 0}} 
&\simeq& -\frac{(1-\beta)}{N} \rho^- \rho_1, \\
\label{Deltarho1-0} 
\Delta \rho_1\big|_{{- \atop 0} \to {- \atop 1}} 
&\simeq& \frac{\beta \,
  \rho_{10}}{2N} \left[ \rho^- - \frac{q}{2}(1-p_d) \rho^{+-} \right]. 
\end{eqnarray}
Adding Eqs.~(\ref{Deltarho1+1}), (\ref{Deltarho1+03}), 
(\ref{Deltarho1-1}) and (\ref{Deltarho1-0}), the rate equation
(\ref{drho1dt}) for $\rho_1$ becomes
\begin{equation}
\frac{d \rho_1}{dt} \simeq \frac{\gamma \beta \, \rho_{10}}{2} 
-(1-\beta) \rho_1,
\label{drho1dt1}
\end{equation}
with 
\begin{equation}
\gamma \equiv 1 - q (1-p_d) \rho^{+-},
\end{equation}
as quoted in Eqs.~(\ref{Drho1Dt}) and (\ref{Gamma}) of the main text.

\section{Derivation of the rate equation for $\rho_{10}$}
\label{rho10}

\begin{figure}[t]
\begin{center}
  \includegraphics[width=7.0cm]{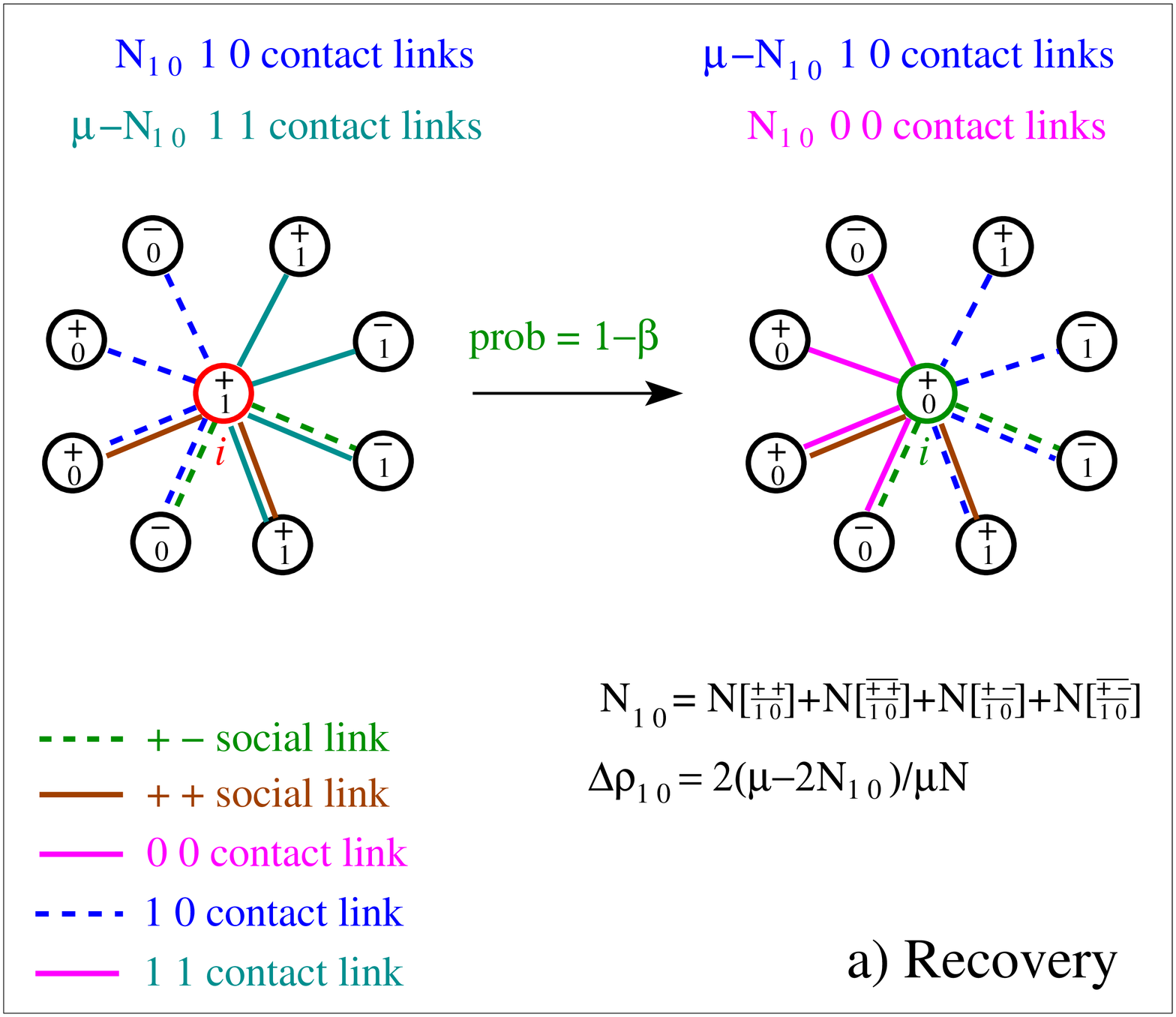}~~~~
  \includegraphics[width=7.0cm]{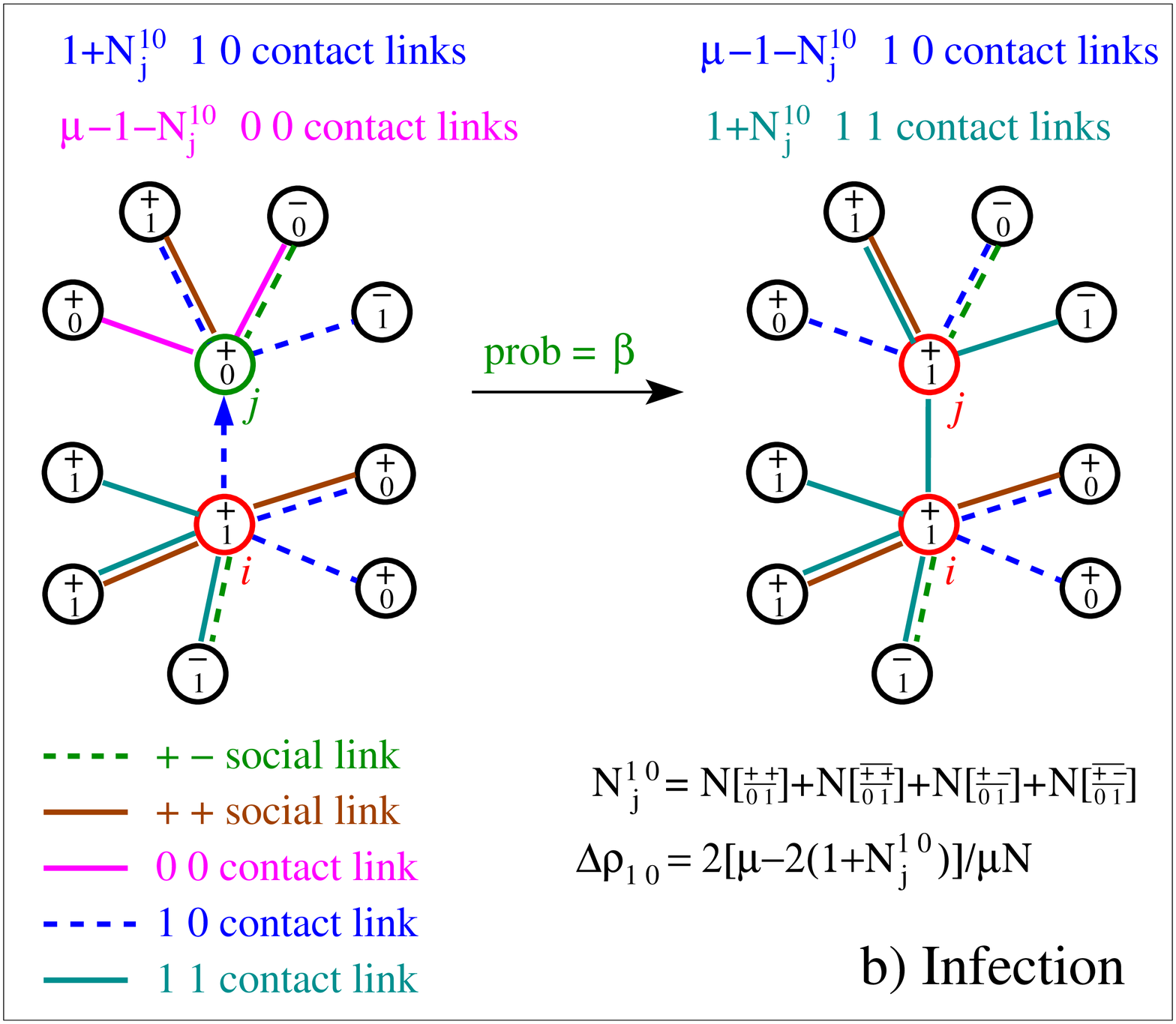}
  \caption{(Color online) Schematic illustration of two disease
    updates.  (a) Recovery: a node $i$ in state $\big[ {+ \atop 1}
      \big]$ recovers with probability 
    $1-\beta$.  (b) Infection: a node $i$ in state 
    $\big[ {+ \atop 1} \big]$ infects a contact neighbor $j$ in 
    state $\big[ {+ \atop 0} \big]$ with probability $\beta$.  The
    change in the density of $10$ contact links is denoted by $\Delta
    \rho_{10}$.} 
\label{contact-update}
\end{center}
\end{figure}

The average change of the fraction of infected-susceptible pairs of
nodes $\rho_{10}$ in a single time step can be written as 
\begin{eqnarray}
\frac{d \rho_{10}}{dt} = \frac{1}{1/N} \left[ 
\Delta \rho_{10}\big|_{{+ \atop 1} \to {+ \atop 0}} +  
\Delta \rho_{10}\big|_{{+ \atop 0} \to {+ \atop 1}} \right] +
\frac{1}{1/N} \left[  
\Delta \rho_{10}\big|_{{+ \atop 1} \to {+ \atop 0}} +  
\Delta \rho_{10}\big|_{{+ \atop 0} \to {+ \atop 1}} \right]^{+ \Leftrightarrow -},    
\label{drho10dt}
\end{eqnarray}
where the first and second terms correspond to the change in
$\rho_{10}$ due to the recovery of a $\big[{+ \atop 1}\big]$ node and
the infection of a $\big[{+ \atop 0}\big]$ node, respectively, while the
last two terms are the corresponding recovery and infections events of
nodes with $-$ opinion, and are obtained by interchanging the symbols
$+$ and $-$ in the first two terms.  The recovery term can be calculated as     
 \begin{eqnarray}
\Delta \rho_{10}\big|_{{+ \atop 1} \to {+ \atop 0}}   
&=& P\left({\mbox{\small $+ \atop 1$}}\right) (1-\beta)
\sum_{\{\mathcal N^+_1 \}}^\mu 
M \left( \big\{\mathcal N^+_1 \big\},\mu \right)
\frac{ \big[\mu -2 \big( 
\mathcal N [\mbox{\tiny ${+ + \atop \overline{1~0}}$}] +  
\mathcal N [\mbox{\tiny ${\overline{++} \atop \overline{1~0}}$}]+
\mathcal N [\mbox{\tiny ${+ - \atop \overline{1~0}}$}] + 
\mathcal N [\mbox{\tiny ${\overline{+-} \atop \overline{1~0}}$}] \big)
\big] }{\mu N/2}, \nonumber \\
\label{Deltarho10+1+0}
\end{eqnarray}
where the expression in square brackets is the change in the number of $10$
links connected to a node $i$ in state $\big[{+ \atop 1}\big]$ when
$i$ recovers, given a specific configuration of node types connected
to $i$ [see Fig.~\ref{contact-update}(a)].  The summation in
Eq.~(\ref{Deltarho10+1+0}) leads to the 
first moments of the multinomial probability 
$M \left( \big\{\mathcal N^+_1 \big\},\mu \right)$, with single event
probabilities 
$P [\mbox{\tiny ${+ + \atop \overline{1~0}}$}]$ and   
$P [\mbox{\tiny ${\overline{++} \atop \overline{1~0}}$}]$
given by Eqs.~(\ref{cond-prob3}), and 
\begin{eqnarray}
P[\mbox{\tiny ${+ - \atop \overline{1 \; 0}}$}] &\simeq& 
\frac{(1-q) \, \rho^- \, \rho_{10}}{2\rho_1}, ~~~
P[\mbox{\tiny ${\overline{+ -} \atop \overline {1 \; 0}}$}] \simeq 
\frac{q \, \rho^{+-} \, \rho_{10}}{4\rho^+ \rho_1}.
\label{cond-prob5}
\end{eqnarray}
Replacing these expressions for the probabilities and using the
conservation relations from Eqs.~(\ref{c11}) and (\ref{c13}) we obtain,
after doing some algebra,  
\begin{eqnarray}
\Delta \rho_{10}\big|_{{+ \atop 1} \to {+ \atop 0}} \simeq 
\frac{2(1-\beta) \rho^+}{N} (\rho_1 - \rho_{10}).
\label{Deltarho10+1+01}
\end{eqnarray} 

We now calculate the second term of Eq.~(\ref{drho10dt})
corresponding to the change in $\rho_{10}$ after the infection of a
node with $+$ opinion:
\begin{eqnarray}
\Delta \rho_{10}\big|_{{+ \atop 0} \to {+ \atop 1}} &=& \Big[ 
P\left({\mbox{\small $+ \atop 1$}}\right) 
\sum_{\{\mathcal N^{~+}_{i,\,1} \}}^\mu M \left( \big\{\mathcal N^{~+}_{i,\,1}
  \big\},\mu \right) \frac{\beta}{\mu} \big( 
\mathcal N_i [\mbox{\tiny ${+ + \atop \overline{1~0}}$}] +  
\mathcal N_i [\mbox{\tiny ${\overline{+ +} \atop \overline{1~0}}$}]
\big) \nonumber \\ &+&
P\left({\mbox{\small $- \atop 1$}}\right) 
\sum_{\{\mathcal N^{~-}_{i,\,1} \}}^\mu M \left( \big\{\mathcal N^{~-}_{i,\,1}
  \big\},\mu \right) \frac{\beta}{\mu} \big( 
\mathcal N_i [\mbox{\tiny ${- + \atop \overline{1~0}}$}] +  
p_d \, \mathcal N_i [\mbox{\tiny ${\overline{- +} \atop \overline{1~0}}$}]
\big) \Big] \nonumber \\ 
&\times&   
\sum_{\{\mathcal N^{~+}_{j,\,0} \}}^{\mu-1} 
M \left( \big\{\mathcal N^{~+}_{j,\,0} \big\},\mu-1 \right)
\frac{ \big[\mu -2 \big( 1 +
\mathcal N_j [\mbox{\tiny ${+ + \atop \overline{0~1}}$}] +  
\mathcal N_j [\mbox{\tiny ${\overline{++} \atop \overline{0~1}}$}]+
\mathcal N_j [\mbox{\tiny ${+ - \atop \overline{0~1}}$}] + 
\mathcal N_j [\mbox{\tiny ${\overline{+-} \atop \overline{0~1}}$}] \big)
\big] }{\mu N/2} \nonumber \\ &=& \mathcal{P \times C}.
\label{Deltarho10+0}
\end{eqnarray}
The term called $\mathcal P$ in Eq.~(\ref{Deltarho10+0}) --the two
summations inside the square brackets-- is the
probability that an $\big[{\mathcal O \atop 1}\big]$ node $i$ infects a
$\big[{+ \atop 0}\big]$ neighbor $j$, and is the same as the one
calculated in Eq.~(\ref{Deltarho1+01}) for    
$\Delta \rho_1 \big|_{{+ \atop 0} \to {+ \atop 1}}$, which is
estimated in Eq.~(\ref{Deltarho1+03}) as  
\begin{equation}
\mathcal P \simeq \frac{\beta \rho_{10}}{2} \left[ \rho^+ -
  \frac{q}{2}(1-p_d) \rho^{+-} \right].  
\label{P} 
\end{equation} 
We notice that the extra $1/N$ prefactor in Eq.~(\ref{Deltarho1+03})
comes from the change in $\rho_1$, which for $\rho_{10}$ depends on the
neighborhood of node $j$.  The subindex $i$ in the term $\mathcal P$
indicates that the infection probability term depends only on node $i$ and 
its neighborhood [see Fig.~\ref{contact-update}(b)].  The term called
$\mathcal C$ 
corresponding to the summation outside the square brackets
expresses the change in $\rho_{10}$ when node $j$ gets infected [see
Fig.~\ref{contact-update}(b)].  Here the subindex $j$ refers to node
$j$ and its neighborhood. 
This term carries the information that the infection on $j$ comes from
one of its infected neighbors $i$, and thus it is known already that
at least one of $j$'s neighbors has disease state $\mathcal D_i=1$.
This is taken into 
account by running the summation on the other $\mu-1$ unknown
neighbors and considering that the number of $10$ links connected
to $j$ is at least one, which is added to the total number of $10$
links inside the parentheses.  Using the conditional probabilities
\begin{eqnarray}
P[\mbox{\tiny ${+ + \atop \overline{0 \; 1}}$}] &\simeq& 
\frac{(1-q) \, \rho^+ \, \rho_{10}}{2\rho_0}, ~~~
P[\mbox{\tiny ${\overline{+ +} \atop \overline {0 \; 1}}$}] \simeq 
\frac{q \, \rho^{++} \, \rho_{10}}{2\rho^+ \rho_0}, \\ 
P[\mbox{\tiny ${+ - \atop \overline{0 \; 1}}$}] &\simeq& 
\frac{(1-q) \, \rho^- \, \rho_{10}}{2\rho_0}, ~~~
P[\mbox{\tiny ${\overline{+ -} \atop \overline {0 \; 1}}$}] \simeq 
\frac{q \, \rho^{+-} \, \rho_{10}}{4\rho^+ \rho_0}, 
\label{cond-prob6}
\end{eqnarray}
and the conservation relations Eqs.~(\ref{c12}) and (\ref{c13}),
the change term becomes
\begin{eqnarray}
\mathcal C \simeq \frac{2}{\mu N} 
\left[ (\mu-1)\left(1-\frac{\rho_{10}}{\rho_0} \right) - 1 \right].
\label{C}
\end{eqnarray}
Finally, combining Eqs.~(\ref{P}) and (\ref{C}) for $\mathcal P$ and
$\mathcal C$ we arrive to
\begin{eqnarray}
\Delta \rho_{10}\big|_{{+ \atop 0} \to {+ \atop 1}} \simeq
\frac{\beta \rho_{10}}{\mu N} \left[ \rho^+ - \frac{q}{2}(1-p_d) \rho^{+-}
  \right] \left[ (\mu-1)\left(1-\frac{\rho_{10}}{\rho_0} \right) - 1 \right].  
\label{Deltarho10+01}
\end{eqnarray}
Then, adding the recovery and infection terms from
Eqs.~(\ref{Deltarho10+1+01}) and (\ref{Deltarho10+01}), respectively,
we obtain the first two terms of Eq.~(\ref{drho10dt}), while the last two terms
are obtained by interchanging symbols $+$ and $-$ in this last
expression.  Adding these four terms we arrive to the following rate
equation for $\rho_{10}$  
\begin{eqnarray}
\frac{d \rho_{10}}{dt} \simeq 
\frac{\gamma \beta \rho_{10}}{\mu} \left[ 
(\mu-1)\left(1-\frac{\rho_{10}}{\rho_0} \right) - 1 \right] + 
2(1-\beta) (\rho_1 - \rho_{10}),  
\end{eqnarray}
quoted in Eq.~(\ref{Drho10Dt}) of the main text.

\bibliographystyle{apsrev}

\bibliography{references}

\begin{thebibliography}{37}
\expandafter\ifx\csname natexlab\endcsname\relax\def\natexlab#1{#1}\fi
\expandafter\ifx\csname bibnamefont\endcsname\relax
  \def\bibnamefont#1{#1}\fi
\expandafter\ifx\csname bibfnamefont\endcsname\relax
  \def\bibfnamefont#1{#1}\fi
\expandafter\ifx\csname citenamefont\endcsname\relax
  \def\citenamefont#1{#1}\fi
\expandafter\ifx\csname url\endcsname\relax
  \def\url#1{\texttt{#1}}\fi
\expandafter\ifx\csname urlprefix\endcsname\relax\def\urlprefix{URL }\fi
\providecommand{\bibinfo}[2]{#2}
\providecommand{\eprint}[2][]{\url{#2}}

\bibitem[{\citenamefont{Barrat et~al.}(2008)\citenamefont{Barrat,
  Barth{\'e}lemy, and Vespignani}}]{Barrat-2008}
\bibinfo{author}{\bibfnamefont{A.}~\bibnamefont{Barrat}},
  \bibinfo{author}{\bibfnamefont{M.}~\bibnamefont{Barth{\'e}lemy}},
  \bibnamefont{and}
  \bibinfo{author}{\bibfnamefont{A.}~\bibnamefont{Vespignani}},
  \emph{\bibinfo{title}{{Dynamical Processes on Complex Networks}}}
  (\bibinfo{publisher}{Cambridge University Press}, \bibinfo{year}{2008}).

\bibitem[{\citenamefont{Castellano et~al.}(2009)\citenamefont{Castellano,
  Fortunato, and Loreto}}]{Castellano-2009}
\bibinfo{author}{\bibfnamefont{C.}~\bibnamefont{Castellano}},
  \bibinfo{author}{\bibfnamefont{S.}~\bibnamefont{Fortunato}},
  \bibnamefont{and} \bibinfo{author}{\bibfnamefont{V.}~\bibnamefont{Loreto}},
  \bibinfo{journal}{Rev. Mod. Phys.} \textbf{\bibinfo{volume}{81}},
  \bibinfo{pages}{591} (\bibinfo{year}{2009}).

\bibitem[{\citenamefont{Pastor-Satorras
  et~al.}(2015)\citenamefont{Pastor-Satorras, Castellano, Van~Mieghem, and
  Vespignani}}]{Pastor-Satorras-2015}
\bibinfo{author}{\bibfnamefont{R.}~\bibnamefont{Pastor-Satorras}},
  \bibinfo{author}{\bibfnamefont{C.}~\bibnamefont{Castellano}},
  \bibinfo{author}{\bibfnamefont{P.}~\bibnamefont{Van~Mieghem}},
  \bibnamefont{and}
  \bibinfo{author}{\bibfnamefont{A.}~\bibnamefont{Vespignani}},
  \bibinfo{journal}{Rev. Mod. Phys.} \textbf{\bibinfo{volume}{87}},
  \bibinfo{pages}{925} (\bibinfo{year}{2015}).

\bibitem[{\citenamefont{Clifford and Sudbury}(1973)}]{Clifford-1973}
\bibinfo{author}{\bibfnamefont{P.}~\bibnamefont{Clifford}} \bibnamefont{and}
  \bibinfo{author}{\bibfnamefont{A.}~\bibnamefont{Sudbury}},
  \bibinfo{journal}{Biometrika} \textbf{\bibinfo{volume}{60}},
  \bibinfo{pages}{581} (\bibinfo{year}{1973}).

\bibitem[{\citenamefont{Holley and Liggett}(1975)}]{Holley-1975}
\bibinfo{author}{\bibfnamefont{R.}~\bibnamefont{Holley}} \bibnamefont{and}
  \bibinfo{author}{\bibfnamefont{T.~M.} \bibnamefont{Liggett}},
  \bibinfo{journal}{Ann. Probab.} \textbf{\bibinfo{volume}{4}},
  \bibinfo{pages}{195} (\bibinfo{year}{1975}).

\bibitem[{\citenamefont{Liggett}(2004)}]{Liggett-2004}
\bibinfo{author}{\bibfnamefont{T.~M.} \bibnamefont{Liggett}},
  \emph{\bibinfo{title}{Interacting Particle Systems}}
  (\bibinfo{publisher}{Springer}, \bibinfo{address}{Berlin ; New York},
  \bibinfo{year}{2004}).

\bibitem[{\citenamefont{Harris}(1974)}]{Harris-1974}
\bibinfo{author}{\bibfnamefont{T.~E.} \bibnamefont{Harris}},
  \bibinfo{journal}{Ann. Prob.} \textbf{\bibinfo{volume}{2}},
  \bibinfo{pages}{969} (\bibinfo{year}{1974}).

\bibitem[{\citenamefont{Marro and Dickman}(1999)}]{Marro-1999}
\bibinfo{author}{\bibfnamefont{J.}~\bibnamefont{Marro}} \bibnamefont{and}
  \bibinfo{author}{\bibfnamefont{R.}~\bibnamefont{Dickman}},
  \emph{\bibinfo{title}{Non-equilibrium Phase Transitions in Lattice Models}}
  (\bibinfo{publisher}{Cambridge University Press}, \bibinfo{year}{1999}).

\bibitem[{\citenamefont{Ferreira and Ferreira}(2013)}]{Ferreira-2013}
\bibinfo{author}{\bibfnamefont{R.}~\bibnamefont{Ferreira}} \bibnamefont{and}
  \bibinfo{author}{\bibfnamefont{S.}~\bibnamefont{Ferreira}},
  \bibinfo{journal}{European Physical Journal B} \textbf{\bibinfo{volume}{86}},
  \bibinfo{pages}{1} (\bibinfo{year}{2013}).

\bibitem[{\citenamefont{De~Domenico et~al.}(2013)\citenamefont{De~Domenico,
  Sole-Ribalta, Cozzo, Kivela, Moreno, Porter, Gomez, and
  Arenas}}]{Domenico-2013}
\bibinfo{author}{\bibfnamefont{M.}~\bibnamefont{De~Domenico}},
  \bibinfo{author}{\bibfnamefont{A.}~\bibnamefont{Sole-Ribalta}},
  \bibinfo{author}{\bibfnamefont{E.}~\bibnamefont{Cozzo}},
  \bibinfo{author}{\bibfnamefont{M.}~\bibnamefont{Kivela}},
  \bibinfo{author}{\bibfnamefont{Y.}~\bibnamefont{Moreno}},
  \bibinfo{author}{\bibfnamefont{M.~A.} \bibnamefont{Porter}},
  \bibinfo{author}{\bibfnamefont{S.}~\bibnamefont{Gomez}}, \bibnamefont{and}
  \bibinfo{author}{\bibfnamefont{A.}~\bibnamefont{Arenas}},
  \bibinfo{journal}{Physical Review X} \textbf{\bibinfo{volume}{3}},
  \bibinfo{pages}{041022} (\bibinfo{year}{2013}).

\bibitem[{\citenamefont{Boccaletti et~al.}(2014)\citenamefont{Boccaletti,
  Bianconi, Criado, Del~Genio, G{\'o}mez-Garde{\~n}es, Romance, Sendiña-Nadal,
  Wang, and Zanin}}]{Boccaletti-2014}
\bibinfo{author}{\bibfnamefont{S.}~\bibnamefont{Boccaletti}},
  \bibinfo{author}{\bibfnamefont{G.}~\bibnamefont{Bianconi}},
  \bibinfo{author}{\bibfnamefont{R.}~\bibnamefont{Criado}},
  \bibinfo{author}{\bibfnamefont{C.~I.} \bibnamefont{Del~Genio}},
  \bibinfo{author}{\bibfnamefont{J.}~\bibnamefont{G{\'o}mez-Garde{\~n}es}},
  \bibinfo{author}{\bibfnamefont{M.}~\bibnamefont{Romance}},
  \bibinfo{author}{\bibfnamefont{I.}~\bibnamefont{Sendiña-Nadal}},
  \bibinfo{author}{\bibfnamefont{Z.}~\bibnamefont{Wang}}, \bibnamefont{and}
  \bibinfo{author}{\bibfnamefont{M.}~\bibnamefont{Zanin}},
  \bibinfo{journal}{Physics Reports} \textbf{\bibinfo{volume}{544}},
  \bibinfo{pages}{1} (\bibinfo{year}{2014}).

\bibitem[{\citenamefont{Kivel{\"a} et~al.}(2014)\citenamefont{Kivel{\"a},
  Arenas, Barth{\'e}lemy, Gleeson, Moreno, and Porter}}]{Kivela-2014}
\bibinfo{author}{\bibfnamefont{M.}~\bibnamefont{Kivel{\"a}}},
  \bibinfo{author}{\bibfnamefont{A.}~\bibnamefont{Arenas}},
  \bibinfo{author}{\bibfnamefont{M.}~\bibnamefont{Barth{\'e}lemy}},
  \bibinfo{author}{\bibfnamefont{J.~P.} \bibnamefont{Gleeson}},
  \bibinfo{author}{\bibfnamefont{Y.}~\bibnamefont{Moreno}}, \bibnamefont{and}
  \bibinfo{author}{\bibfnamefont{M.~A.} \bibnamefont{Porter}},
  \bibinfo{journal}{Journal of Complex Networks} \textbf{\bibinfo{volume}{2}},
  \bibinfo{pages}{203} (\bibinfo{year}{2014}).

\bibitem[{\citenamefont{Granell et~al.}(2013)\citenamefont{Granell, G\'omez,
  and Arenas}}]{Granell-2013}
\bibinfo{author}{\bibfnamefont{C.}~\bibnamefont{Granell}},
  \bibinfo{author}{\bibfnamefont{S.}~\bibnamefont{G\'omez}}, \bibnamefont{and}
  \bibinfo{author}{\bibfnamefont{A.}~\bibnamefont{Arenas}},
  \bibinfo{journal}{Phys. Rev. Lett.} \textbf{\bibinfo{volume}{111}},
  \bibinfo{pages}{128701} (\bibinfo{year}{2013}).

\bibitem[{\citenamefont{Granell et~al.}(2014)\citenamefont{Granell, G{\'o}mez,
  and Arenas}}]{Granell-2014}
\bibinfo{author}{\bibfnamefont{C.}~\bibnamefont{Granell}},
  \bibinfo{author}{\bibfnamefont{S.}~\bibnamefont{G{\'o}mez}},
  \bibnamefont{and} \bibinfo{author}{\bibfnamefont{A.}~\bibnamefont{Arenas}},
  \bibinfo{journal}{Physical Review E} \textbf{\bibinfo{volume}{90}},
  \bibinfo{pages}{012808} (\bibinfo{year}{2014}).

\bibitem[{\citenamefont{Alvarez-Zuzek et~al.}(2016)\citenamefont{Alvarez-Zuzek,
  La~Rocca, Vazquez, and Braunstein}}]{Zuzek-2016}
\bibinfo{author}{\bibfnamefont{L.~G.} \bibnamefont{Alvarez-Zuzek}},
  \bibinfo{author}{\bibfnamefont{C.~E.} \bibnamefont{La~Rocca}},
  \bibinfo{author}{\bibfnamefont{F.}~\bibnamefont{Vazquez}}, \bibnamefont{and}
  \bibinfo{author}{\bibfnamefont{L.~A.} \bibnamefont{Braunstein}},
  \bibinfo{journal}{PLOS ONE} \textbf{\bibinfo{volume}{11}},
  \bibinfo{pages}{e0163593} (\bibinfo{year}{2016}).

\bibitem[{\citenamefont{Halu et~al.}(2013)\citenamefont{Halu, Zhao,
  Baronchelli, and Bianconi}}]{Halu-2013}
\bibinfo{author}{\bibfnamefont{A.}~\bibnamefont{Halu}},
  \bibinfo{author}{\bibfnamefont{K.}~\bibnamefont{Zhao}},
  \bibinfo{author}{\bibfnamefont{A.}~\bibnamefont{Baronchelli}},
  \bibnamefont{and} \bibinfo{author}{\bibfnamefont{G.}~\bibnamefont{Bianconi}},
  \bibinfo{journal}{Eurphys. Lett.} \textbf{\bibinfo{volume}{102}},
  \bibinfo{pages}{16002} (\bibinfo{year}{2013}).

\bibitem[{\citenamefont{Diakonova et~al.}(2014)\citenamefont{Diakonova,
  San~Miguel, and Egu{\'i}luz}}]{Diakonova-2014}
\bibinfo{author}{\bibfnamefont{M.}~\bibnamefont{Diakonova}},
  \bibinfo{author}{\bibfnamefont{M.}~\bibnamefont{San~Miguel}},
  \bibnamefont{and} \bibinfo{author}{\bibfnamefont{V.~M.}
  \bibnamefont{Egu{\'i}luz}}, \bibinfo{journal}{Phys. Rev. E}
  \textbf{\bibinfo{volume}{89}}, \bibinfo{pages}{062818}
  (\bibinfo{year}{2014}).

\bibitem[{\citenamefont{Diakonova et~al.}(2016)\citenamefont{Diakonova,
  Nicosia, Latora, and San~Miguel}}]{Diakonova-2016}
\bibinfo{author}{\bibfnamefont{M.}~\bibnamefont{Diakonova}},
  \bibinfo{author}{\bibfnamefont{V.}~\bibnamefont{Nicosia}},
  \bibinfo{author}{\bibfnamefont{V.}~\bibnamefont{Latora}}, \bibnamefont{and}
  \bibinfo{author}{\bibfnamefont{M.}~\bibnamefont{San~Miguel}},
  \bibinfo{journal}{New Journal of Physics} \textbf{\bibinfo{volume}{18}},
  \bibinfo{pages}{023010} (\bibinfo{year}{2016}).

\bibitem[{\citenamefont{Vazquez et~al.}(2016)\citenamefont{Vazquez, Serrano,
  and Miguel}}]{Vazquez-2016}
\bibinfo{author}{\bibfnamefont{F.}~\bibnamefont{Vazquez}},
  \bibinfo{author}{\bibfnamefont{M.~A.} \bibnamefont{Serrano}},
  \bibnamefont{and} \bibinfo{author}{\bibfnamefont{M.~S.}
  \bibnamefont{Miguel}}, \bibinfo{journal}{Scientific Reports}
  \textbf{\bibinfo{volume}{6}}, \bibinfo{pages}{29342} (\bibinfo{year}{2016}).

\bibitem[{\citenamefont{Czaplicka et~al.}(2016)\citenamefont{Czaplicka, Toral,
  and San~Miguel}}]{Czaplicka-2016}
\bibinfo{author}{\bibfnamefont{A.}~\bibnamefont{Czaplicka}},
  \bibinfo{author}{\bibfnamefont{R.}~\bibnamefont{Toral}}, \bibnamefont{and}
  \bibinfo{author}{\bibfnamefont{M.}~\bibnamefont{San~Miguel}},
  \bibinfo{journal}{Phys. Rev. E} \textbf{\bibinfo{volume}{94}},
  \bibinfo{pages}{062301} (\bibinfo{year}{2016}).

\bibitem[{\citenamefont{Axelrod}(1997)}]{Axelrod-1998}
\bibinfo{author}{\bibfnamefont{R.}~\bibnamefont{Axelrod}},
  \bibinfo{journal}{Journal of Conflict Resolution}
  \textbf{\bibinfo{volume}{41}}, \bibinfo{pages}{203} (\bibinfo{year}{1997}).

\bibitem[{\citenamefont{McPherson et~al.}(2001)\citenamefont{McPherson,
  Smith-Lovin, and Cook}}]{McPherson-2001}
\bibinfo{author}{\bibfnamefont{M.}~\bibnamefont{McPherson}},
  \bibinfo{author}{\bibfnamefont{L.}~\bibnamefont{Smith-Lovin}},
  \bibnamefont{and} \bibinfo{author}{\bibfnamefont{J.~M.} \bibnamefont{Cook}},
  \bibinfo{journal}{Annual Review of Sociology} \textbf{\bibinfo{volume}{27}},
  \bibinfo{pages}{415} (\bibinfo{year}{2001}).

\bibitem[{\citenamefont{Vazquez and Redner}(2007)}]{Vazquez-2007-2}
\bibinfo{author}{\bibfnamefont{F.}~\bibnamefont{Vazquez}} \bibnamefont{and}
  \bibinfo{author}{\bibfnamefont{S.}~\bibnamefont{Redner}},
  \bibinfo{journal}{Eurphys. Lett.} \textbf{\bibinfo{volume}{78}},
  \bibinfo{pages}{18002} (\bibinfo{year}{2007}).

\bibitem[{\citenamefont{Castellano et~al.}(2003)\citenamefont{Castellano,
  Vilone, and Vespignani}}]{Castellano-2003}
\bibinfo{author}{\bibfnamefont{C.}~\bibnamefont{Castellano}},
  \bibinfo{author}{\bibfnamefont{D.}~\bibnamefont{Vilone}}, \bibnamefont{and}
  \bibinfo{author}{\bibfnamefont{A.}~\bibnamefont{Vespignani}},
  \bibinfo{journal}{Europhys. Lett.} \textbf{\bibinfo{volume}{63}},
  \bibinfo{pages}{153} (\bibinfo{year}{2003}).

\bibitem[{\citenamefont{Vilone and Castellano}(2004)}]{Vilone-2004}
\bibinfo{author}{\bibfnamefont{D.}~\bibnamefont{Vilone}} \bibnamefont{and}
  \bibinfo{author}{\bibfnamefont{C.}~\bibnamefont{Castellano}},
  \bibinfo{journal}{Phys. Rev. E} \textbf{\bibinfo{volume}{69}},
  \bibinfo{pages}{016109} (\bibinfo{year}{2004}).

\bibitem[{\citenamefont{Suchecki et~al.}(2005)\citenamefont{Suchecki,
  Egu\'{\i}luz, and San~Miguel}}]{Suchecki-2005}
\bibinfo{author}{\bibfnamefont{K.}~\bibnamefont{Suchecki}},
  \bibinfo{author}{\bibfnamefont{V.~M.} \bibnamefont{Egu\'{\i}luz}},
  \bibnamefont{and}
  \bibinfo{author}{\bibfnamefont{M.}~\bibnamefont{San~Miguel}},
  \bibinfo{journal}{Europhys. Lett.} \textbf{\bibinfo{volume}{69}},
  \bibinfo{pages}{228} (\bibinfo{year}{2005}).

\bibitem[{\citenamefont{Castellano et~al.}(2005)\citenamefont{Castellano,
  Loreto, Barrat, Cecconi, and Parisi}}]{Castellano-2005}
\bibinfo{author}{\bibfnamefont{C.}~\bibnamefont{Castellano}},
  \bibinfo{author}{\bibfnamefont{V.}~\bibnamefont{Loreto}},
  \bibinfo{author}{\bibfnamefont{A.}~\bibnamefont{Barrat}},
  \bibinfo{author}{\bibfnamefont{F.}~\bibnamefont{Cecconi}}, \bibnamefont{and}
  \bibinfo{author}{\bibfnamefont{D.}~\bibnamefont{Parisi}},
  \bibinfo{journal}{Phys. Rev. E} \textbf{\bibinfo{volume}{71}},
  \bibinfo{pages}{066107} (\bibinfo{year}{2005}).

\bibitem[{\citenamefont{Sood and Redner}(2005)}]{Sood-2005}
\bibinfo{author}{\bibfnamefont{V.}~\bibnamefont{Sood}} \bibnamefont{and}
  \bibinfo{author}{\bibfnamefont{S.}~\bibnamefont{Redner}},
  \bibinfo{journal}{Physical Review Letters} \textbf{\bibinfo{volume}{94}},
  \bibinfo{pages}{178701} (\bibinfo{year}{2005}).

\bibitem[{\citenamefont{Vazquez and Egu{\'i}luz}(2008)}]{Vazquez-2008-2}
\bibinfo{author}{\bibfnamefont{F.}~\bibnamefont{Vazquez}} \bibnamefont{and}
  \bibinfo{author}{\bibfnamefont{V.~M.} \bibnamefont{Egu{\'i}luz}},
  \bibinfo{journal}{New Journal of Physics} \textbf{\bibinfo{volume}{10}},
  \bibinfo{pages}{063011} (\bibinfo{year}{2008}).

\bibitem[{\citenamefont{Castellano et~al.}(2000)\citenamefont{Castellano,
  Marsili, and Vespignani}}]{Castellano-2000}
\bibinfo{author}{\bibfnamefont{C.}~\bibnamefont{Castellano}},
  \bibinfo{author}{\bibfnamefont{M.}~\bibnamefont{Marsili}}, \bibnamefont{and}
  \bibinfo{author}{\bibfnamefont{A.}~\bibnamefont{Vespignani}},
  \bibinfo{journal}{Phys. Rev. Lett.} \textbf{\bibinfo{volume}{85}},
  \bibinfo{pages}{3536} (\bibinfo{year}{2000}).

\bibitem[{\citenamefont{Parshani et~al.}(2010)\citenamefont{Parshani, Buldyrev,
  and Havlin}}]{Parshani-2010}
\bibinfo{author}{\bibfnamefont{R.}~\bibnamefont{Parshani}},
  \bibinfo{author}{\bibfnamefont{S.~V.} \bibnamefont{Buldyrev}},
  \bibnamefont{and} \bibinfo{author}{\bibfnamefont{S.}~\bibnamefont{Havlin}},
  \bibinfo{journal}{Phys. Rev. Lett.} \textbf{\bibinfo{volume}{105}},
  \bibinfo{pages}{048701} (\bibinfo{year}{2010}).

\bibitem[{\citenamefont{Hu et~al.}(2011)\citenamefont{Hu, Ksherim, Cohen, and
  Havlin}}]{Hu-2011}
\bibinfo{author}{\bibfnamefont{Y.}~\bibnamefont{Hu}},
  \bibinfo{author}{\bibfnamefont{B.}~\bibnamefont{Ksherim}},
  \bibinfo{author}{\bibfnamefont{R.}~\bibnamefont{Cohen}}, \bibnamefont{and}
  \bibinfo{author}{\bibfnamefont{S.}~\bibnamefont{Havlin}},
  \bibinfo{journal}{Phys. Rev. E} \textbf{\bibinfo{volume}{84}},
  \bibinfo{pages}{066116} (\bibinfo{year}{2011}).

\bibitem[{\citenamefont{Gao et~al.}(2012)\citenamefont{Gao, Buldyrev, Stanley,
  and Havlin}}]{Gao-2012}
\bibinfo{author}{\bibfnamefont{J.}~\bibnamefont{Gao}},
  \bibinfo{author}{\bibfnamefont{S.~V.} \bibnamefont{Buldyrev}},
  \bibinfo{author}{\bibfnamefont{H.~E.} \bibnamefont{Stanley}},
  \bibnamefont{and} \bibinfo{author}{\bibfnamefont{S.}~\bibnamefont{Havlin}},
  \bibinfo{journal}{Nature Physics} \textbf{\bibinfo{volume}{8}},
  \bibinfo{pages}{40} (\bibinfo{year}{2012}).

\bibitem[{\citenamefont{Zhou et~al.}(2014)\citenamefont{Zhou, Bashan, Cohen,
  Berezin, Shnerb, and Havlin}}]{Zhou-2014}
\bibinfo{author}{\bibfnamefont{D.}~\bibnamefont{Zhou}},
  \bibinfo{author}{\bibfnamefont{A.}~\bibnamefont{Bashan}},
  \bibinfo{author}{\bibfnamefont{R.}~\bibnamefont{Cohen}},
  \bibinfo{author}{\bibfnamefont{Y.}~\bibnamefont{Berezin}},
  \bibinfo{author}{\bibfnamefont{N.}~\bibnamefont{Shnerb}}, \bibnamefont{and}
  \bibinfo{author}{\bibfnamefont{S.}~\bibnamefont{Havlin}},
  \bibinfo{journal}{Phys. Rev. E} \textbf{\bibinfo{volume}{90}},
  \bibinfo{pages}{012803} (\bibinfo{year}{2014}).

\bibitem[{\citenamefont{Radicchi and Arenas}(2013)}]{Radicchi-2013}
\bibinfo{author}{\bibfnamefont{F.}~\bibnamefont{Radicchi}} \bibnamefont{and}
  \bibinfo{author}{\bibfnamefont{A.}~\bibnamefont{Arenas}},
  \bibinfo{journal}{Nature Physics} \textbf{\bibinfo{volume}{9}},
  \bibinfo{pages}{717} (\bibinfo{year}{2013}).

\bibitem[{\citenamefont{Radicchi}(2014)}]{Radicchi-2014}
\bibinfo{author}{\bibfnamefont{F.}~\bibnamefont{Radicchi}},
  \bibinfo{journal}{Phys. Rev. X} \textbf{\bibinfo{volume}{4}},
  \bibinfo{pages}{021014} (\bibinfo{year}{2014}).

\bibitem[{\citenamefont{Zhao and Bianconi}(2013)}]{Zhao-2013}
\bibinfo{author}{\bibfnamefont{K.}~\bibnamefont{Zhao}} \bibnamefont{and}
  \bibinfo{author}{\bibfnamefont{G.}~\bibnamefont{Bianconi}},
  \bibinfo{journal}{Journal of Statistical Physics}
  \textbf{\bibinfo{volume}{152(6)}}, \bibinfo{pages}{1069}
  (\bibinfo{year}{2013}).

\end{thebibliography}

\end{document}